\documentclass[runningheads,oribiblxs]{llncs}
\usepackage[T1]{fontenc}
\usepackage{graphicx}
\usepackage[sectionbib]{natbib}
\renewcommand{\bibfont}{\footnotesize}

\usepackage{comment}
\usepackage{xspace}
\usepackage{bm}
\usepackage{bbm}
\usepackage{tikz}
\usetikzlibrary{trees}
\usetikzlibrary{quotes}
\usepackage{nicefrac}

\usepackage[font=small,labelfont=bf]{caption}
\usepackage{subcaption}
\usepackage{amsmath}
\usepackage{amssymb}
\usepackage[bibliography=common]{ apxproof}

\renewcommand{\bibname}{References}

\usepackage{float}
\usepackage{color}
\usepackage{listings}
\usepackage{setspace}
\usepackage{algorithmicx}
\usepackage[Algorithm,ruled]{algorithm}
\usepackage{algpseudocode}
\usepackage{etoolbox}

\patchcmd{\thebibliography}{\chapter*}{\section*}{}{}

\setcitestyle{citesep={;}}

\DeclareMathOperator*{\argmax}{arg\,max}

\newcommand{\abs}[1]{\lvert #1 \rvert}

\newtheoremrep{theorem}{Theorem}
\newtheoremrep{proposition}{Proposition}
\newtheoremrep{definition}{Definition}

\newcommand{\tg}{\textsc{Game$_1$}\xspace}
\newcommand{\otg}{\textsc{Game$_2$}\xspace}
\newcommand{\fourRound}{\textsc{Game$_3$}\xspace}
\newcommand{\NF}{\mathit{NF}}
\newcommand{\TE}{\mathit{TE}}

\newcommand{\child}{\mathsf{child}}
\newcommand{\parent}{\mathsf{parent}}
\newcommand{\cI}{\mathcal{I}}
\newcommand{\bsigma}{\bm{\sigma}}
\newcommand{\bbR}{\mathbb{R}}
\newcommand{\regret}{\mathsf{Reg}}
\newcommand{\expect}{\mathbb{E}}
\newcommand{\Binom}{\mathsf{Binom}}
\newcommand{\Prob}{\mathsf{Pr}}
\newcommand{\cA}{\mathit{2A}}
\newcommand{\cB}{\mathit{2B}}
\newcommand{\cC}{\mathit{1C}}
\newcommand{\cD}{\mathit{1D}}

\usepackage{newfloat}
\usepackage{listings}
\floatstyle{ruled}
\newfloat{listing}{tb}{lst}{}
\floatname{listing}{Listing}

\begin{document}
\title{Exploiting Extensive-Form Structure in Empirical Game-Theoretic Analysis}
%
%
\author{Christine Konicki \and
Mithun Chakraborty \and
Michael P. Wellman}
\authorrunning{C. Konicki et al.}
%
\institute{University of Michigan, Ann Arbor MI 48109, USA \\
\email{\{ckonicki,dcsmc,wellman\}@umich.edu}}
\maketitle              
\begin{abstract}
Empirical game-theoretic analysis (EGTA) is a general framework for reasoning about complex games using agent-based simulation.
Data from simulating select strategy profiles is employed to estimate a cogent and tractable game model approximating the underlying game.
To date, EGTA methodology has focused on game models in normal form; though the simulations play out in sequential observations and decisions over time, the game model abstracts away this temporal structure.
Richer models of \textit{extensive-form games} (EFGs) provide a means to capture temporal patterns in action and information, using tree representations. 
We propose \textit{tree-exploiting EGTA} (TE-EGTA), an approach to incorporate EFG models into EGTA\@.
TE-EGTA constructs game models that express observations and temporal organization of activity, albeit at a coarser grain than the underlying agent-based simulation model.
The idea is to exploit key structure while maintaining tractability.
We establish theoretically and experimentally that exploiting even a little temporal structure can vastly reduce estimation error in strategy-profile payoffs compared to the normal-form model.%
\footnote{This paper has been slightly revised from the original version published at WINE 2022; to wit, the proof included in the appendices of our key theoretical result has been expanded.}
Further, we explore the implications of EFG models for iterative approaches to EGTA, where strategy spaces are extended incrementally.
Our experiments on several game instances demonstrate that TE-EGTA can also improve performance in the iterative setting, as measured by the quality of equilibrium approximation as the strategy spaces are expanded.

\end{abstract}

\section{Introduction}\label{sec:intro}
\textit{Empirical game-theoretic analysis} (EGTA) \citep{Wellman16} employs agent-based simulation to induce a game model over a restricted set of strategies. 
The methodology is salient for games that are too complex for analytic description and reasoning.
Complexity in dynamics and information can be expressed in a simulator, but abstracted from the game model.
In typical EGTA practice, simulation data is used to estimate a \textit{normal-form game} (NFG) model, associating a payoff vector with each combination of strategies available to the agents. But game theory offers richer model forms that capture sequentiality in agent play and conditional information.
Specifically, \textit{extensive-form game} (EFG) models represent the game as a tree, where nodes or sets of nodes represent states, and edges represent player moves and chance events. 
Whereas NFGs treat agent strategies as atomic objects, EFGs afford a finer-grained expression of the observations and actions that define these strategies, capturing structure that may be shared among many strategies. 
The goal of this work is to take advantage of extensive-form structure, at flexible granularity, for complex game environments described by agent-based simulation. 
Our approach, \textit{Tree-Exploiting EGTA} (TE-EGTA), follows the basic framework of EGTA, but employs a parameterized EFG model to leverage part of the game's tree structure.

Taking advantage of extensive form necessitates two key modifications to the EGTA process.
First, we require methods to estimate the more complex model form: an abstracted game tree parameterized by player utilities at terminal nodes and probability distributions over successors for stochastic events represented by \textit{chance nodes} in the tree. 
These stochastic events, together with \textit{information-set} structure, model the \textit{imperfect information} available to the players. 
We introduce straightforward techniques to estimate these game-tree parameters, and describe how the structure intuitively affords more effective use of available simulation data.
Second, we require methods for extending extensive-form models as the strategy space is expanded, across iterations of the EGTA process.
We introduce techniques for iterative augmentation of empirical game-tree models with new (best-response) strategies, within a standard approach that incorporates deep RL within EGTA \citep{psro17}. 

To establish the benefits of tree-exploitation for EGTA, we show that an extensive-form empirical game model provides (with high probability) a more accurate approximation of the true game than a normal-form model constructed from the same simulation data.
As it is generally intractable to construct a game tree expressing the full fidelity of the game simulated, our approach is designed to operate on highly abstracted models capturing only selected tree structure.
To ground the meaning of such abstractions, we provide an algorithm that produces a coarsened model given the full game and a description of what to abstract away. 
We demonstrate the efficacy of TE-EGTA through experiments on three stylized games, and over varying levels of abstraction.
We compare TE-EGTA to normal-form EGTA on two key performance measures.
The first is the average error incurred from estimating the true player payoffs for all strategy combinations in the empirical game.
The second is the \textit{regret} of empirical-game solutions with respect to the full multiagent scenario, computed over successive empirical game models in an iterative EGTA process.

\textit{Outline.} 
\S\ref{sec:prelim} provides technical preliminaries, including a formal exposition of the EFG representation and precise elaboration of the EGTA framework and process.
\S\ref{sec:tree_expl_egta} delineates our algorithmic contribution, TE-EGTA, starting with the structure of an extensive-form empirical game model and how to estimate its parameters from simulation data (\S\ref{sec:TE_EGTA_param_est}). We then give a theoretical procedure for generating a (usually) coarsened extensive-form model from the underlying game (\S\ref{sec:abstraction}), and explain how to  iteratively refine the model via simulation-aided strategy exploration (\S\ref{sec:treePSRO}). 
In \S\ref{sec:theo_improv}, we present theoretical results on the advantage of TE-EGTA over normal-form EGTA in approximating true payoffs given a set of strategy profiles. All proofs are available in the full version.
In \S\ref{sec:expts}, we report experiments that demonstrate the improvement in strategy-profile payoff estimation (\S\ref{sec:exp1}) and in model refinement using the PSRO approach \citep{psro17} (\S\ref{sec:exp2}) produced via tree exploitation. 
\S\ref{sec:disc} concludes.

\section{Preliminaries}\label{sec:prelim}
\subsection{Extensive-Form Games (EFGs)}\label{sec:EFG_model}

An \textit{extensive-form game} (EFG) is a standard model for strategic multi-agent scenarios where agents act \textit{sequentially} with potentially varying degrees of \textit{imperfect information} about the history of game play. 
Early algorithmic work on EFGs showed how to generalize the Lemke-Howson method for computing Nash equilibria (NE) for two-player games with perfect recall \citep{koller96}. 
Well-known game-theoretic methods such as replicator dynamics \citep{gatti13} and fictitious self-play \citep{fsp_15} have also been adapted for EFGs. The task of successful abstraction with exploitability guarantees has also been investigated:
\cite{kroer18} gave a framework for analyzing abstractions of large-scale EFGs, and  \cite{certificates20} introduced the notion of small certificates carrying proofs of approximate NE\@. 
Other works have developed algorithms that search for optimal strategies or approximate equilibria that minimize exploitability \citep{johanson12,lockhart19}. In this paper, we will only consider games with perfect recall, so no player can forget what it observed or knew earlier.

\begin{figure}[ht!]
	\centering
	\includegraphics[scale=0.32]{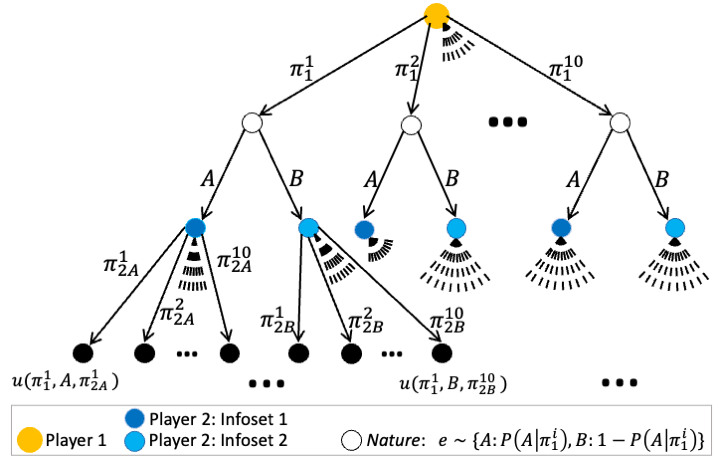}
	\caption{EFG representation of \tg, our running example also used in our experiments. Dashed lines indicate outgoing edges to nodes omitted from this illustration.}
	\label{fig:game1}
\end{figure}

\textbf{Tree structure.} Formally, a finite, imperfect-information EFG is a tuple $G:=\langle N, H, V, \{\mathcal{I}_j\}_{j=0}^n, \{\Pi_j\}_{j=1}^n, X, P, u \rangle$. The components of $G$ are defined as follows (see Fig.~\ref{fig:game1} for an illustrative example):  
\begin{itemize}
    \item $N = \{0, \dotsc, n\}$ is the set of \textit{players}. Player~0 represents \textit{Nature}, a non-strategic agent responsible for stochastic events that impact the course of play; the remaining players are strategic rational agents.
    \item $H$ is the finite game tree, rooted at a node $h_0$, that captures the dynamic nature of interactions. Each node $h \in H$ represents a \textit{state} of the game, also identified with a history of actions (see below) beginning at the \textit{initial state} $h_0$ which corresponds to the null history $\emptyset$.
    The leaves or \textit{terminal nodes} $T \subset H$ represent possible end-states of the game. We refer to the non-terminal nodes of $H$ as \textit{decision nodes}, represented by the set $D = H \setminus T$.
    \item $V: D \rightarrow N$ assigns a player to each decision node $h$.
    \item For each player $j \in N$, $\cI_j$ is a partition of $V^{-1}(j)$ where each $I \in \cI_j$ is an \textit{information set (infoset)} of $j$. All nodes $h \in I$ are indistinguishable from the viewpoint of player $j$.
    \item At each information set $I \in \cI_j$, player $j$ has a set of available actions $\Pi_j(I)$\@.
    \item A node $h$ where $V(h) = 0$ is called a \textit{chance node}. $X(h)$ is the set of actions available to Nature (i.e., possible outcomes of the stochastic event) at $h$, and $P(\cdot \mid h)$ is the probability distribution over $X(h)$.
    \item The \textit{utility function} $u: T \rightarrow \bbR^n$ maps each terminal node to a real-valued vector of players' utilities $\{u_j(t)\}_{j=1}^n$.
\end{itemize}

The directed edge connecting any $h \in I$ to its child $\child[h]$ represents a state transition resulting from $V(h)$'s move, and is labeled with an action $\pi \in \Pi_{V(h)}(I)$ if $V(h) \neq 0$, or an outcome $x \in X(h)$ otherwise.
We denote by $\varphi(h, j)$ the history of actions belonging to player $j$ up to node $h$.

\textbf{Strategies and payoffs.}  
A \textit{pure strategy} for player $j \in N \setminus \{0\}$ specifies the 
action $\pi_j \in \Pi_j(I)$ that $j$ selects at information set $I \in \cI_j$.
More generally, a \textit{mixed strategy} or simply \textit{strategy} $\sigma_j(\cdot \mid I)$ defines a probability distribution over $\Pi_j(I)$ at each information set of agent $j$; that is, action $\pi_j$ is selected with probability $\sigma_j(\pi_j \mid I)$.
The vector $\bsigma = (\sigma_1, \dotsc, \sigma_n)$ is called a \textit{strategy profile}, and $\bsigma_{-j}$ represents the combination of strategies for players other than $j$. 
We denote the set of all strategies available to player $j$ by~$\Sigma_j$ and the space of joint strategy profiles by $\Sigma= \times_{j=1}^n \Sigma_j$. 
Let $r_j(t, \sigma_j)$ denote the probability that node $t$ is reached if player $j$ adopts strategy $\sigma_j$ and all other players (including Nature) always choose actions that lead to $h$ when possible; the probability that $t$ is reached under strategy profile $\bsigma$ is given by its \textit{reach probability}, $r(t, \bsigma) = \prod_{j \in N} r_j(t, \sigma_j)$. Likewise, the contribution of Nature to the reach probability of $t$ is $r_0(t) = \prod_{h \in H, \, e \in X(h) \, \cap \, \varphi(t, 0)} P\left( e \mid h \right)$. We define the \textit{payoff} from joint strategy profile $\bsigma$ to player $j$ as its expected utility over all end-states: $U_j(\bsigma) :=  \sum_{t \in T} u_j(t) r(t, \bsigma)$.

\textbf{Best response formulation and regret.}
A \textit{best response (BR)} of player $j \in N \setminus \{0\}$ to $\bsigma_{-j}$ is a strategy $\sigma_j \in \argmax_{\sigma'_j \in \Sigma_j} U_j(\sigma'_j, \bsigma_{-j})$ that maximizes the payoff for $j$ given $\bsigma_{-j}$. The \textit{regret} of player~$j$ from playing $\bsigma$ is given by
$\regret_j(\bsigma) = \max_{\sigma_j \in \Sigma_j} U_j(\sigma_j, \bsigma_{-j}) - U_j(\bsigma)$.
The total regret of the strategy profile $\bsigma$ is the sum: $\regret(\bsigma) = \sum_{j=1}^n \regret_j(\bsigma)$.
For $\varepsilon > 0$, an $\varepsilon$-\textit{Nash equilibrium} is a strategy profile $\bsigma$ such that $\regret_j(\bsigma) \le \varepsilon$ for every player $j \in N \setminus \{0\}$; a strategy profile $\bsigma$ with $\regret(\bsigma) = 0$ is a \textit{Nash equilibrium}.


\textbf{Running example.} Consider the two-agent strategic scenario depicted in Fig.~\ref{fig:game1}, which we call \tg. 
First, Player~1  chooses an action from $\Pi_1 = \{ \pi^i_1 \}_{i=1}^{10}$; then, a single stochastic event $X(\pi^i_1) \in \{A, B\}$ occurs, outcome $A$ having probability $P(A \mid \pi^i_1)$ dependent on Player~1's choice $\pi^i_1$. 
 Player~2 observes the outcome $e\in\{A,B\}$ but not Player~1's chosen action, which induces two information sets for Player~2. Player~2 also has ten actions to choose from in each information set, $\Pi_{\cA} = \{ \pi^i_{\cA} \}_{i=1}^{10}$ and $\Pi_{\cB} = \{ \pi^i_{\cB} \}_{i=1}^{10}$. 
 Each leaf with history $(\pi^i_1, e, \pi^{i'}_{\textit{2e}})$ is labeled with the $2$-dimensional vector of Player~1 and~2's realized  utilities. 
 Neither the conditional probabilities $P(A \mid \pi^i_1)$ nor the leaf utilities $u(\pi^i_1, e, \pi^{i'}_{\textit{2e}})$ are known \textit{a priori} to the game analyst.


\subsection{Empirical Game-Theoretic Analysis (EGTA)}\label{sec:NF_EGTA}

The framework of EGTA was developed for the application of game-theoretic reasoning to scenarios too complex for analytic description, accessible only in the form of a procedural simulation \citep{Wellman16}.
Over the years, EGTA has been applied to multifarious problem domains including recreational strategy games \citep{Tuyls20}, security games \citep{wang19sywsjf}, social dilemmas \citep{Leibo17}, and auctions \citep{Wellman20}.
There is also substantial work on methodological questions such as how to decide which strategy profiles to simulate \citep{Fearnley15,Jordan08vw}, and how to reason statistically about estimated game models \citep{viqueria19,Tuyls20,Vorobeychik10}.
Recently, EGTA has received newfound attention, as the simulation-based approach meshes well with powerful new strategy generation methods from deep reinforcement learning (RL)  \citep{psro17}.

\begin{figure}[ht!]
	\centering
	\includegraphics[scale=0.28]{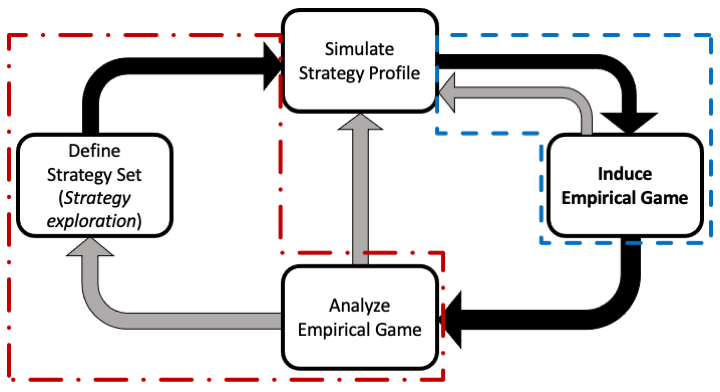}
	\caption{Schematic illustration of EGTA. TE-EGTA modifies two subprocesses to incorporate the tree structure of EFGs: accumulation of simulation data into the game model (enclosed in blue, described in \S\ref{sec:TE_EGTA_param_est}); and the procedure for augmenting $\hat{\Sigma}$ with new strategies (enclosed in red, described in \S\ref{sec:treePSRO}). Black (resp. grey) arrows represent the sequence of operations (resp. direction of possible information flow).}
	\label{fig:egtaBlock}
\end{figure}

The main feature of EGTA is its construction of an \textit{empirical game model} $\hat{G}$ of a much larger game of interest, called the \textit{true game} $G$, from simulation data. A typical EGTA process (see Fig.~\ref{fig:egtaBlock}) iteratively refines and extends $\hat{G}$ by cumulative simulation over an incrementally growing strategy space. 
$\hat{G}$ is a \textit{simplification} of the underlying $G$ since: (1) it is defined on restricted subsets $\hat{\Sigma}_j\subset {\Sigma}_j$ of the players' true-game strategy spaces, and 
 the \textit{restricted strategy profile space}, given by $\hat{\Sigma}= \times_{j=1}^n \hat{\Sigma}_j$, is typically a vast reduction of $\Sigma$; (2) some information revelation  and conditioning structure may be abstracted away.
Moreover, we assume that $G$ is accessible only through a high-fidelity but expensive \textit{simulator} that executes a given strategy profile in $G$ and outputs limited observation histories and noisy utility samples. $\hat{G}$ is thus also an \textit{approximation} of $G$ since its parameters must be estimated from this simulation data.
 
 
 Almost all EGTA literature to date expresses game models in \textit{normal form}, given by a (multi-dimensional) matrix of payoff estimates for combinations of agents' strategies from the restricted set. 
 The multi-agent scenarios themselves are typically dynamic in nature, as represented by an agent-based simulator; agent strategies are generally conditional on partial observations. For example, a normal-form game model for \tg in \S\ref{sec:EFG_model} would treat each pure strategy $\pi_2^{i'}$ of player 2 as atomic, abstracting away the nuanced conditioning on whether $A$ or $B$ happened, and record estimated utility vectors for strategy combinations of the form $(\pi_1^i,\pi_2^{i'})$ from restricted set.
 

 As our objective is to extend EGTA to extensive-form modeling, we will call this normal-form baseline \textit{NF-EGTA}.
 In NF-EGTA, the sole simulator output of concern is the noisy sample of players' payoffs, from which we compute estimates $\{ \hat{U}^{\NF}_j(\bsigma)\}_{j = 1}^n$ of the true utilities $\{U_j(\bsigma)\}_{j=1}^n$ to obtain the empirical game model $\hat{G}$.
 We then analyze or solve this tractable, multi-dimensional game matrix by standard techniques to obtain a result for the next iteration.
 Termination may be decided by a criterion such as the \textit{true-game regret} of a solution (i.e., the maximum payoff increase achievable by any player $j$ by deviating to a strategy in $\Sigma_j$ rather than $\hat{\Sigma}_j$) falling below a specified threshold.
If termination criteria are not met, we expand the restricted strategy sets through a process called \textit{strategy exploration} \citep{Balduzzi19,wellman10}, and update $\hat{G}$ through further simulation and model induction. 



\textbf{Game Model Estimation.} Consider the process of estimating a normal-form model for an underlying extensive-form game implicitly represented by traces from the simulator.
Suppose we simulate each strategy profile in $\hat{\Sigma}$ $m$ times. 
Each simulated play traces a path through the game tree ending at some undisclosed terminal node $t \in T$ and returns a vector of noisy payoffs for all players sampled from a distribution with expectation $u(t)$. Let $\{ \bar{u}^i_j \}_{j = 1}^n$ denote the realized payoff sample at the end of the $i^\mathrm{th}$ simulation for $i = 1,\dotsc,m$;
Typically, NF-EGTA's payoff estimate $\hat{U}^{\NF}_j(\bsigma)$ is the simple average of these samples.
 $\hat{U}^{\NF}_j(\bsigma)$ is an unbiased estimator of the true payoff, as shown in Proposition~\ref{prop:unbiased_NF}. In practice, the number of samples $m$ that can be acquired is limited by the computational cost of simulation. This begs the question: can incorporating tree structure into $\hat{G}$ improve the accuracy of estimated payoffs, relative to \textit{NF-EGTA}, for a fixed simulation budget $m$? We address this question in \S\ref{sec:TE_EGTA_param_est}.
\begin{proposition}\label{prop:unbiased_NF}
For every player $j \in N \setminus \{0\}$ and strategy profile $\bsigma \in \hat{\Sigma}$, $\expect\left[ \hat{U}^{\NF}_j(\bsigma) \right] = U_j(\bsigma)$. 
\end{proposition}

\textbf{Policy-Space Response Oracles (PSRO).} A fully automated implementation of the iterative EGTA framework of Fig.~\ref{fig:egtaBlock} requires the ability to automatically generate new strategies based on analysis of the empirical game model at a given point.
\cite{Phelps06} first introduced automated strategy generation to EGTA via genetic search, and \cite{egta_rl09} first employed RL for this purpose.
The advent of deep RL methods brought significant new power to this approach, which is now the predominant means of accumulating a set of restricted strategies in EGTA algorithms.

\cite{psro17} developed a general framework for interleaving empirical game modeling with deep RL techniques, which they termed \textit{policy-space response oracles}.
A key idea of PSRO is that of a \textit{meta-strategy solver} (MSS), an abstract operation that implements the ``Analyze Empirical Game'' block of Fig.~\ref{fig:egtaBlock}.
The output of an MSS is a strategy profile, which provides the other-agent context for a BR calculation performed by deep RL\@. 
The policy generated by RL as a BR to the MSS result is then added as a new strategy to expand the current restricted strategy space, leading to another round of simulation and induction for the next EGTA iteration. The MSS concept provides a useful abstraction for expressing a variety of approaches to strategy exploration \citep{Wang22mw}.
For example, using Nash equilibrium as an MSS yields the double oracle (DO) algorithm \citep{do_03}.
If the MSS simply returns the uniform distribution over the restricted strategy sets, the algorithm reduces to fictitious play.

Prior work has extended the DO algorithm to exploit game-tree structure.
\cite{bosansky2014exact} developed a \textit{sequence-form double-oracle} algorithm for zero-sum EFGs that maintains a restricted game model based on partial action sequences.
The XDO algorithm of \cite{McAleer21} for two-player zero-sum games computes a mixed BR at each information set, as compared to normal-form DO which mixes policies only at the root level. It modifies PSRO for EFGs while still using a normal-form empirical model.
The benefits over normal-form demonstrated by these works suggest EGTA can be similarly extended to exploit game-tree structure beyond the strategy exploration block.


\section{Tree-Exploiting EGTA}\label{sec:tree_expl_egta}
We call our approach for augmenting empirical game models to incorporate extensive-form game elements \textit{tree-exploiting EGTA} (TE-EGTA).
In the typical normal-form treatment of EGTA, the underlying game is parameterized by entries in a payoff matrix $\{U_j(\bsigma)\}_{j \in N \setminus \{0\}, \bsigma \in \Sigma}$.%
\footnote{More general approaches based on regression have been proposed \citep{Sokota19,Vorobeychik07}, which also amount to parameterized representations of a payoff function.}
TE-EGTA instead parameterizes the underlying game to capture the EFG tree structure through a set of \textit{leaf utilities} $\{u(t)\}_{t \in T}$, 
and \textit{conditional probability distributions} that are dependent on possibly unobserved previous choices made in the game play and estimated from observations of stochastic events. 

We assume that the structure of decisions and stochastic events in the empirical EFG model is given (typically a high-level abstraction of the game tree implicitly represented by the simulator, as discussed in \S\ref{sec:abstraction}).
This ensures that the order of player choices and stochastic events in the empirical game tree matches the order in the true game, from root to leaf. 
In particular, the true game's information sets must be a refinement of the empirical game's information sets.
Given this structure, we treat observations of Nature's actions as conditioned on past game play.
The empirical game tree therefore must associate with each chance node a conditional probability distribution over the relevant set of outgoing edges.
Leaves of the tree are associated with payoff estimates, which depend on the entire path from the root.

Each simulation of a strategy profile yields sample payoffs, as well as a trace of publicly or privately \textit{observable actions} from both the players and Nature that are made over the course of the game.
This is a key point of contrast with the normal-form model, for which only payoffs are relevant.
The trace of actions tells us which leaf node in the abstract model is reached and what stochastic event outcomes were realized along the way.

To explain our tree-exploiting estimation approach, we first restate the expression for $U_j(\bsigma)$ 
 in a way that explicitly factors in probabilities of specific \textit{observations} of stochastic events. We assume that a game theorist working with the black-box simulator's partial observations in order to formulate an empirical model is aware of the game's rules, and so can surmise where in the game the observation has occurred. We also assume that the observation labels used by the simulator allow the game theorist to distinguish the observations from each other and associate them with the appropriate chance nodes. 
 A stochastic observation during gameplay 
 is captured in the tree by an edge $e \in \varphi(t, 0)$ from a chance node $h$ such that $V(h) = 0$ to a node with history $he$. 
The reach probability of $he$ from the perspective of Nature is $r_0(he) = P(e \mid h)$, and recall $r_0(t)$ is the joint probability of Nature's choices along the path from the root to $t$. 
Hence, 
\begin{align}
U_j(\bsigma) &= \sum_{t \in T} u_j(t) \prod_{k = 1}^n r_k(t, \sigma_k) r_0(t).  \label{eq:te_util_exp}
 \end{align}
 
 \subsection{TE-EGTA Game Model Estimation}\label{sec:TE_EGTA_param_est}

The probabilities $r_k(t, \sigma_k)$, for all terminal nodes~$t$, are directly determined by the strategy profile $\bsigma$. 
Hence, to estimate $U_j(\bsigma)$ based on Eq.~\eqref{eq:te_util_exp}, we need estimates for $u(t)$ and $\{r_0(t)\}_{t \in T}$\@.
These are, in fact, the game parameters for TE-EGTA (leaf utilities and conditional probabilities respectively) that we introduced above. We denote the respective estimates by $\{\hat{u}_j(t)\}_{j=1}^n$ and $\{\hat{r}_0(t)\}_{t \in T}$.
 
A key feature of TE-EGTA is that, in modeling the payoff of strategy profile~$\bsigma$, we estimate the parameters using \textit{all} relevant simulation data, not just the data from simulating $\bsigma$. 
Different strategy profiles may lead to overlapping or identical paths being taken through the game tree, with some probability.
 We compute $\hat{u}_j(t)$ as the sample average of player~$j$'s payoffs across simulation runs that terminate at node $t$.
Similarly, we estimate chance node probabilities using all simulations. Suppose a chance node $h$ is reached $m_h$ times across all simulation data, and the node with history $he$ (reflecting Nature's choice $e$) is reached $m_{he} < m_h$ times.
 The empirical probability of observing the stochastic outcome represented by $e$ in the game tree is $\frac{m_{he}}{m_h}$. 
Note that $m_h$ can never be zero because the algorithm for constructing the empirical game model includes only nodes that are reached in simulation.
Finally, we give player~$j$'s estimated payoff for strategy profile~$\bsigma$:
\begin{align*}
    \hat{U}^{\TE}_j(\bsigma) 
    &= \sum_{t \in T} \hat{u}_j(t) \prod_{k = 1}^n r_k(t, \sigma_k) \left( \prod_{e \in \varphi(t, 0)} \frac{m_{he}}{m_{h}} \right).
\end{align*}
Recall that each strategy profile $\bsigma$ in $\hat{\Sigma}$ is simulated $m$ times, resulting in $m$ game play sequences for each. 
Some strategies that end at different terminal nodes $t_1$ and $t_2$ may still include the same node $h$ in their respective paths and result in the same observation $e \in \hat{X}(h)$. The observation occurs with the same probability for both strategies since their histories diverge only at node $h e$. This feature is what allows the empirical game model to take into account the role of different decision points in the formulation of player strategies in a way that the normal-form model does not. 

To illustrate the difference in model estimation between NF- and TE-EGTA, consider the following example from \tg.
Suppose we simulate the strategy profile $(\pi_1^1,\pi_2^1)$ 10 times, and obtain the following payoff samples for Player~1: $99,95,100,96,95,100,92,95,93,94$; 
we also observe outcome $A$ of the stochastic event in the first $6$ of these $10$ simulations. 
NF-EGTA would simply average the 10 payoff samples and record $\hat{U}^{\NF}_1(\pi_1^1,\pi_2^1)=95.9$.  In contrast,
TE-EGTA distinguishes the 6 samples corresponding to the leaf $(\pi^1_1, A, \pi^1_{\cA})$ from the 4 samples corresponding to the leaf $(\pi^1_1, B, \pi^1_{\cB})$, and separately averages them to get the estimates $\hat{u}_1(\pi^1_1, A, \pi^1_{\cA})=97.5$ and $\hat{u}_1(\pi^1_1, B, \pi^1_{\cB})=93.5$. 
Now, suppose we also have data from $10$ simulations of another strategy profile $(\pi_1^1,\pi_2^2)$, $\pi_2^2 \neq \pi_2^1$, 
$A$ being realized in $5$ of these simulations. 
From this experience, our overall estimated probability of $A$ conditioned on $\pi_1^1$ is $\tfrac{6+5}{10+10}=0.55$. 
Thus, using all relevant sample data, $\hat{U}^{\TE}_1(\pi_1^1,\pi_2^1) = 0.55\times 97.5+(1-0.55)\times 93.5=95.7$.

The following proposition shows that, like NF-EGTA, TE-EGTA produces unbiased estimates of strategy-profile payoffs. However, our theoretical results in \S\ref{sec:theo_improv} suggest that TE-EGTA offers more accurate payoff estimates with a high probability.
\begin{proposition}\label{prop:teegta_est}
 For every player $j \in N \setminus \{0\}$ and strategy profile $\bsigma \in \hat{\Sigma}$, $\expect_{t \sim r(T, \bsigma)} \left[ \hat{U}^{\TE}_j(\bsigma) \right] = U_j(\bsigma)$. 
\end{proposition}

\subsection{The Game Model as an Abstraction}\label{sec:abstraction}
Abstraction methods have extended the state of the art in solving imperfect-information games over the years \citep{sandholm10}, particularly poker. An abstraction algorithm takes as input a complete game description and produces a simpler version of the tree. TE-EGTA incorporates some of the tree structure from the true game into the empirical game model; in order to ground this game model as a coarse abstraction of the underlying game, we describe \textbf{Coarsen}, an algorithm that coarsens a game tree by abstracting away chance nodes. 


We express \textit{coarseness} as the fraction of chance nodes from the true game that are included in the empirical game model. 
An empirical game that matches the true game's structure would include all of them; conversely, an empirical game in normal-form would include none of them. 
We are primarily concerned with games represented by agent-based simulation where the representation of the true game as an EFG is intractable, and thus we would not expect to obtain a coarsened model by actually applying \textbf{Coarsen}.
Our intent is to contextualize a coarsened game as one that could in principle be produced by abstracting away chance nodes.

\begin{algorithm}
\renewcommand{\thealgorithm}{}
\small
\floatname{algorithm}{Coarsen}
\caption{Algorithm for coarsening an input game $G$}
\label{alg:coarsen}
\begin{algorithmic}
\Require{Input game $G$, partition $C' \subseteq C$ and map $\rho: C' \rightarrow X$}
\State Copy $H' = H$, with each node $h$ represented by its history

\For{$c \in C'$, beginning at the chance node furthest from the root}
\State Let $I_j(c)$ be the set of infosets induced by each event $e \in X(c)$ for player $j$.

\State Compute power set $Z^{*}$ of intersections $Z = \bigcap_{I \in I_j(c)} \{ h \mid he \in I \}$ of all the histories $h$ across $I_j(c)$.
\For{$Z \in Z^{*}$}
\State $\{ I'_j, \Pi'_j(I'_j) \}, H' = \textbf{CoarsenInfosets}\left(I_j(c), Z, \rho, G \right)$

\State Assign $X'(c) = X(c) \setminus \rho(c)$
\State $\mathcal{I}'_j, \Pi'_j = \textbf{CondenseBranching} \left( \{ I'_j, \Pi'_j(I'_j) \}, \mathcal{I}'_j \right)$
\EndFor
\EndFor

\State Assign $X'(c) = X(c)$ for all $c \notin C'$.
\State Assign all player $j$'s infosets not conditioned on chance events from any $c \in C'$ to $\mathcal{I}'_j$
\State For all nodes $h$ that preceded or did not follow any nodes in $C'$, assign $V'(h) = V(h)$\\
\vspace{1em}
\Return $G' = (N, H', V', \{\mathcal{I}'\}_{j = 1}^n, \{ \Pi'_j \}_{j = 1}^n, X')$
\end{algorithmic}
\end{algorithm}

The algorithm is given a partition of both $G$'s chance nodes $C = \{ h \in H \mid V(h) = 0\}$ and the set of outcomes $X(h)$ for each chance node, denoting what to exclude from the coarsened tree. One important restriction on $G$ is that the child nodes of a given chance node in $C'$ must all belong to the same player so that they can be collapsed into one node.
We denote the abstracted game by $G' = \langle N, H', V', \{\mathcal{I}'_j\}_{j=1}^n,\{ \Pi'_j\}_{j=1}^n, X' \rangle$ whose components are defined as in~\S\ref{sec:prelim}. 
The nodes identified, information sets, and action spaces will necessarily differ from those of $G$, depending on what information is coarsened and where. Without loss of generality, \textbf{Coarsen} treats both $G$ and $G'$ as binary trees in order to limit the branching factor of $G'$. \textbf{CoarsenInfosets} transforms the intersecting information sets of the children of each $c \in C'$ into a new information set for $G'$ whose action space is the Cartesian product of the old infosets' action spaces. To keep the branching factor equal to 2, \textbf{CondenseBranching} transforms these action spaces (comprised of tuples) into binary (sub-)trees where each edge is part of an action tuple.

\begin{algorithm}
\renewcommand{\thealgorithm}{}
\small
\floatname{algorithm}{CoarsenInfosets}
\caption{Subroutine for coarsening input game G's infosets}
\begin{algorithmic}
\Require{Set $I_j(c)$ of infosets induced by each outcome $e \in X(c)$ for player $j$, set $Z$ of intersecting histories across $I_j(c)$, map $\rho: C' \rightarrow X$, input game $G$, $H'$}
\vspace{1em}
\State Create a new infoset $I'_j = \bigcup_{e \in \rho(c)} \{ he \hspace{0.2em} | \hspace{0.2em} h \in Z \}$ and add to $\mathcal{I}'_j$
\State Compute the new action space $\Pi'_j(I'_j) = \bigotimes_{I \in I_j(c)} \Pi_j(I)$.
\For{$ha \in H'$}
\If{$a$ was part of an action space of $I_j(c)$}
\State Let $\Pi'_j(I'_j, a) = \{ x | x \in \Pi'_j(I'_j), a \in x \}$ 
\State Replace $ha$ with $hb$ for each action tuple $b \in \Pi'_j(I'_j, a)$
\State Assign $V'(hb) = V(ha)$
\EndIf
\EndFor
\State In $\{ I'_j \}$ and $H'$, delete both duplicate histories and from each history, all $e \in \rho(C')$

\vspace{1em}
\Return $\{ I'_j, \Pi'_j(I'_j) \}$, $H'$
\end{algorithmic}
\end{algorithm}

\begin{algorithm}
\renewcommand{\thealgorithm}{}
\small
\floatname{algorithm}{CondenseBranching}
\caption{Subroutine for reducing the branching factor induced by the newly coarsened action tuples}
\begin{algorithmic}
\Require{Set of infosets and action spaces $\{ I'_j, \Pi'_j(I'_j) \}$, final output set $\mathcal{I}'_j$}
\vspace{1em}
\State $A = \mathsf{copy}(\Pi'_j(I'_j))$
\For{$h \in I'_j$}
\State $g = \mathsf{copy}(h)$ and $\Pi'_j(\{ g \}) = \langle \rangle$
\For{$x \in A, a \in x$}
\State Add new infoset $\{ g \}$ to $\mathcal{I}'_j$ if $\{ g \} \notin \mathcal{I}'_j$
\State Add $x[:\mathsf{index}(a)]$ to the action space $\Pi'_j(\{ g \})$ if $a \notin \Pi'_j(\{ g \})$
\State $g = g \hspace{0.2em} x[:\mathsf{index}(a)]$
\State $V'(g) = j$
\EndFor
\EndFor\\
\vspace{1em}
\Return $\mathcal{I}'_j$, $\Pi'_j$
\end{algorithmic}
\end{algorithm}

\subsection{Tree-Exploiting PSRO}\label{sec:treePSRO}
Recall the PSRO framework for iterative EGTA with deep RL, introduced in~\S\ref{sec:NF_EGTA}. 
Like EGTA more generally, past work within the PSRO framework has relied on normal-form representations of the empirical game, even though the games of interest are inherently sequential.
We call PSRO that uses a normal-form (resp. tree-exploiting) empirical game NF-PSRO (TE-PSRO).
In addition to exploiting extensive structure for estimation (\S\ref{sec:TE_EGTA_param_est}), TE-PSRO also takes advantage of the tree representation for managing the restricted strategy space.
A single pure strategy profile can result in multiple different paths depending on Nature's choices. If a new best response for a given infoset is part of the profile, new paths with their own new utilities and stochastic distributions at Nature's decision points are discovered and added to the empirical game tree. If one of those paths includes moves from other players that are already part of the game tree, then additional samples from this new combination can be included in the (tighter) estimation of the old parameters pertinent to that path.

\begin{figure}[ht!]
    \begin{subfigure}[b]{0.42\textwidth}
    \centering
      \includegraphics[scale=0.28]{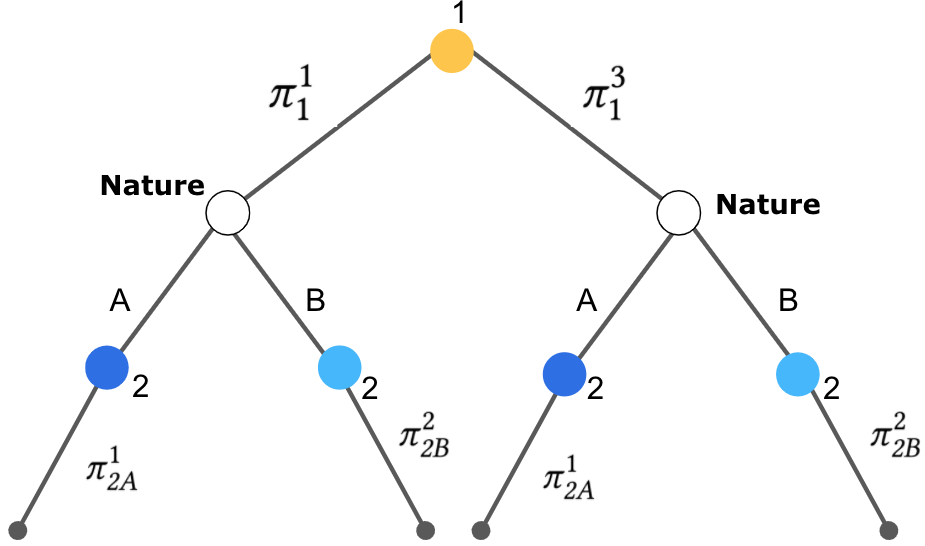}
    \caption{Starting empirical game tree. \label{fig:psro_orig}}
    \end{subfigure}~
    \begin{subfigure}[b]{0.49\textwidth}
     \centering
     \includegraphics[scale=0.21]{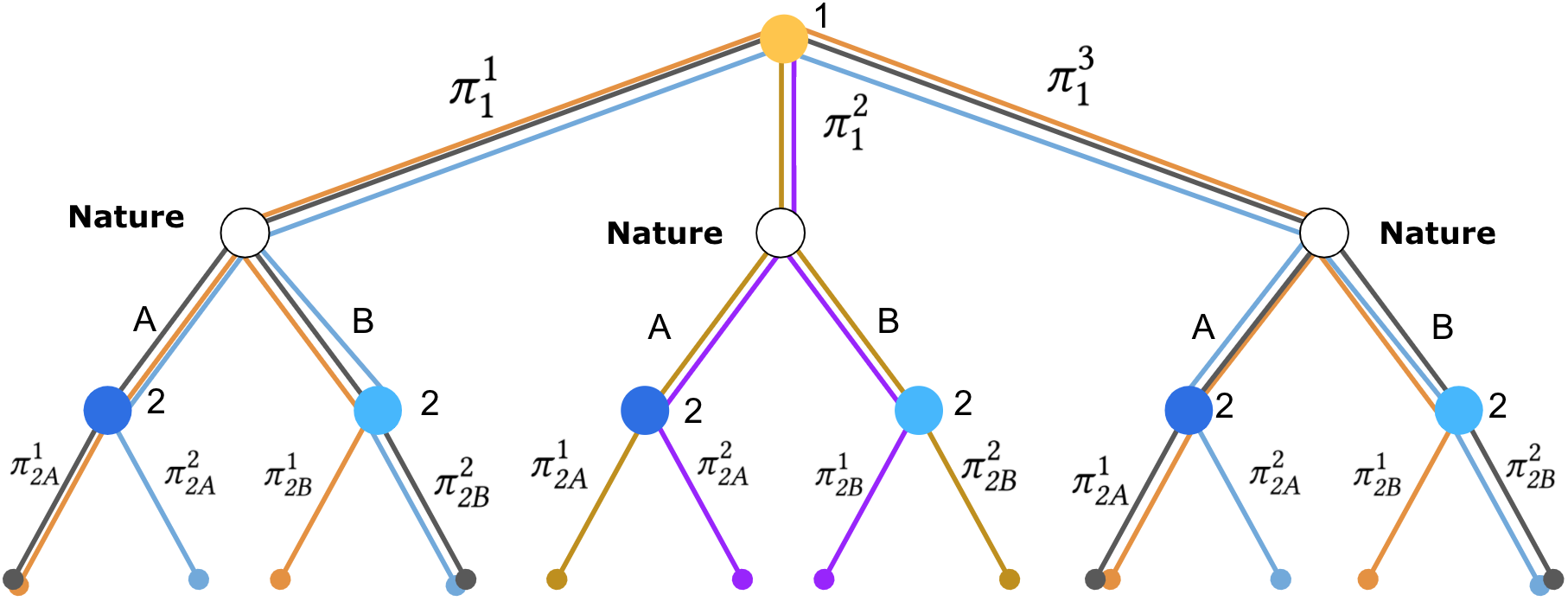}
    \caption{Update after best-response computation. \label{fig:psro_mod}}
    \end{subfigure}
    \caption{Two successive steps of possible TE-PSRO instantiation on \tg. \label{fig:psro}}
\end{figure}

Consider the empirical game in Fig.~\ref{fig:psro_orig} with restricted strategy sets $\hat{\Pi}_1, \hat{\Pi}_{\cA}$, and $\hat{\Pi}_{{\cB}}$ for each information set as shown; the true game here is \tg. 
Let $BR_1(\sigma_{\cA}, \sigma_{{\cB}})$ and $(BR_{\cA}(\sigma_1), BR_{{\cB}}(\sigma_1))$ denote the respective best responses from \tg (the true game) to the strategy profile $(\sigma_1, (\sigma_{\cA}, \sigma_{{\cB}}))$. Suppose, in an iteration, $BR_1(\sigma_{\cA}, \sigma_{{\cB}}) = \pi^2_1$,  $BR_{\cA}(\sigma_1) = \pi_{\cA}^2$, and $BR_{{\cB}}(\sigma_1) = \pi_{{\cB}}^1$.

In the next round, the new best-response elements are considered in conjunction with the pre-existing strategy combinations from the restricted set, as well as other players' new best responses. 
The resulting trajectories are shown in Fig.~\ref{fig:psro_mod}: (1) $BR_1 \times \Pi_{\cA} \times \Pi_{{\cB}}$ highlighted in yellow; (2) $\Pi_1 \times BR_{\cA} \times \Pi_{{\cB}}$ highlighted in blue; (3) $\Pi_1 \times \Pi_{\cA} \times BR_{{\cB}}$ highlighted in orange; and (4) $(BR_1, BR_{\cA}, BR_{{\cB}})$ highlighted in purple. See the full paper for more detail.
This expansion of the empirical game tree captures finer-grained structural information about the true game than simply adding a matrix entry for each new best-response combination. To conclude this section, we supply the pseudocode that summarizes TE-PSRO.

\begin{algorithm}
\label{alg:te-psro}
\renewcommand{\thealgorithm}{}
\small
\floatname{algorithm}{TE-PSRO}
\caption{Tree-Exploiting Policy Space Response Oracles}
\begin{algorithmic}

\Require{Initial singleton strategy sets $\hat{\Sigma}_j$ for all players}
\vspace{1em}
\State Initialize solution profile to the pure strategy $\sigma_j \in \hat{\Sigma}_j$

\While{epoch $e$ in $\{1, 2, \cdots\}$}
\For{player $j \in \{1, 2, \ldots, n\}$}
\State Initialize $Q^{\pi}_j(I, a) = 0$ for all reachable infosets and actions defined by $\bm{\sigma}$
\For{many episodes}
\State Initialize $I$ to the singleton infoset containing the root node
\State Sample $\pi_{-j} \sim \bm{\sigma}_{-j}$
\State Simulate gameplay in $G$ until a leaf is reached using $\pi_{-j}$ and $Q^{\pi}_j(I, a)$,\\ choosing actions $a \in \pi_j(I)$ randomly with prob. $\epsilon$
\State Update $Q$-table with episode rewards
\State Update $\pi'_j(I) \in \argmax_{a} Q^{\pi}_j(I, a)$ for all $I \in \mathcal{I}_j$ included in $\hat{G}$
\EndFor
\State $\hat{\Sigma}_j = \hat{\Sigma}_j \cup \{ \, \pi'_j \, \}$
\EndFor
\State Accumulate new payoff and observation data from black-box simulator
\State Update $\hat{G}$'s average leaf utilities $\hat{u}_j(t)$ with new payoff samples and leaves
\State Update $\hat{G}$'s stochastic probs $\hat{r}_0(t)$ with new observations and chance nodes
\State Compute $\bm{\sigma}$ from $\hat{G}$ and $\hat{\Sigma}$ using an MSS
\EndWhile

\vspace{1em}
\Return A final solution strategy $\sigma_j$ for each player $j$
\end{algorithmic}
\end{algorithm}

\section{Payoff Estimation Improvement: Theoretical Results}\label{sec:theo_improv}
To develop a formal framework for comparing the efficacy of payoff estimation (\S\ref{sec:TE_EGTA_param_est}) by TE-EGTA and NF-EGTA, we apply the concept of \textit{uniform approximation of a game} \citep{viqueria19} to our setting. 
Consider a true EFG $G$ and an empirical game $\hat{G}$  with the same set of players and with restricted set $\hat{\Sigma}$ constructed from accumulated simulation data upon termination of EGTA. Let $\hat{U}_j(\bsigma)$ be the estimate in $\hat{G}$ of an arbitrary player $j$'s true payoff under strategy profile $\bsigma$.
\begin{definition}\label{def:uniform_approx}
The $\ell_{\infty}$-\textit{norm} between games $G$ and $\hat{G}$ is given by
\begin{equation*}
    \parallel G - \hat{G}\parallel_{\infty} = \max_{j \in N \setminus \{0\}, \bsigma \in \hat{\Sigma}} \abs{U_j(\bsigma) - \hat{U}_j(\bsigma)}.
\end{equation*}
If $\parallel G - \hat{G}\parallel_{\infty} \leq \varepsilon$, then $\hat{G}$ is said to be a \textit{uniform $\varepsilon$-approximation} of $G$.
\end{definition}
Note that in this definition, the maximization is only over the restricted set $\hat{\Sigma} \subseteq \Sigma$. 
An important consequence of $\hat{G}$ being a uniform approximation of $G$ upon EGTA's termination is that a strategy profile that is an approximate Nash equilibrium in $\hat{G}$ is an approximate Nash equilibrium in $G$ as well:
\begin{proposition}\label{prop:uni_approx_reg}
If $\hat{G}$ is a uniform $\varepsilon$-approximation of $G$ and $\bsigma$ is a $\gamma$-Nash equilibrium of $\hat{G}$ for some $\gamma \ge 0$, then $\regret_j(\bsigma) \le 2\varepsilon + \gamma$ for each player $j \in N \setminus \{0\}$ upon the termination of EGTA.
\end{proposition}

The main result of this section is that for a given EFG, under reasonable assumptions, TE-EGTA induces an empirical game model that is a tighter uniform approximation of the EFG than that induced by NF-EGTA, with a high probability. Given an arbitrary true game $G$, let $\hat{G}_{\NF}$ and $\hat{G}_{\TE}$ denote respectively the empirical game models induced by the application of NF-EGTA and TE-EGTA to $G$ over the same restricted set $\hat{\Sigma}$.%
\footnote{In the iterative application of EGTA, the NF- and TE- variants may produce different choices of strategies to add; hence, strategy sets covered at a given iteration number tend to diverge. However,
for comparing model estimation accuracy, however, it makes sense to start with a common baseline of strategy space. 
Our experiments (\S\ref{sec:expts}) provide empirical corroboration that the benefits accrue as well when we examine the trajectory of models produced within the iterative PSRO framework.}
We further assume an upper and a lower bound for each agent payoff sample returned by the simulator; more specifically, we assume that the noise function for each payoff follows a sub-Gaussian distribution.
Let $c$ be the number of strategy profiles from the restricted set that, after each profile is sampled $m$ times, result in a path taken through the tree that includes the first edge of $\varphi(t)$. $c$ can be as small as~1 and as large as $O(\abs{\Sigma_j})$ for some $j \in N$ depending on the game structure and when the selected EGTA method terminates. With very high probability, $c$ is strictly greater than ~1 by the time EGTA has terminated due to our exhaustive simulation of all possible profiles in the empirical strategy space. Combined with Proposition~\ref{prop:uni_approx_reg}, we have the following result, which also implies a tighter upper bound for player regret in $G$ under approximate equilibria in the empirical game model computed using payoffs estimated through TE-EGTA.
\begin{theorem}\label{thm:TE_tighter}
For any $\delta \in (0,1)$ and the same number $m$ of game simulation repetitions in each iteration of either type of EGTA, there exist positive constants $\varepsilon_{\NF}$ and $\varepsilon_{\TE}$ such that 
$\frac{\varepsilon_{\TE}}{\varepsilon_{\NF}} = \frac{1}{\sqrt{c}}$,
and with probability at least $1 - \delta$ w.r.t. the randomness in the simulator payoff output, $\hat{G}_{\NF}$ (respectively, $\hat{G}_{\TE}$) is a uniform $\varepsilon_{\NF}$-approximation (respectively, $\varepsilon_{\TE}$-approximation) of $G$.
\end{theorem}

\section{Experiments} \label{sec:expts}
We conducted two sets of experiments comparing TE-EGTA with varying levels of tree structure exploitation to NF-EGTA. Each set used three different EFGs, chosen so that the corresponding empirical game models induced by our flexible tree-exploiting framework would vary in size and complexity. We implemented a simulator for each game that produced observations in accordance with the corresponding stochastic events, and end-state payoff samples that were normally distributed about the true utilities at the respective terminal nodes with a noise variance $\epsilon = 0.1$. The first game was \tg (\S\ref{sec:EFG_model}). In our experiments, for each instance of \tg, we randomly assigned $P(A \mid \pi^i_1)$ from $U[0,1]$ for each $\pi^i_1 \in \Pi_1$ and $u(t)$ from $\{0, 0.25,\dotsc, 4.75, 5\}$ for each leaf utility. During each game play sample, the simulator returned the realized outcome $A$ or $B$ of the single stochastic event and a noisy payoff vector.

\begin{figure}[ht!]
	\centering
	\includegraphics[scale=0.35]{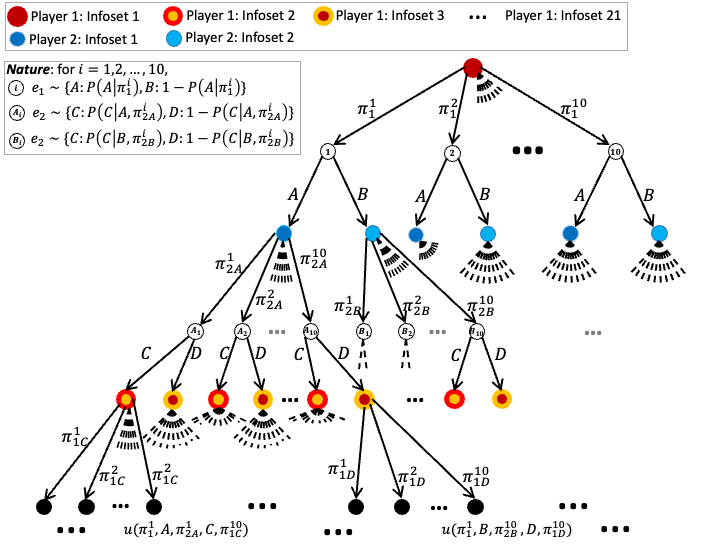}
	\caption{EFG representation of \otg. Dashed lines indicate edges to nodes omitted from this illustration.}
	\label{fig:game2}
\end{figure}

The second game was \otg, an extension of \tg having a second stochastic event $e_2 \in \{ C, D\}$ after Player 2's turn and a second turn for Player 1 afterward. Player 1 only observes its first action and the second event $e_2$. Thus Player 2 has $2$ information sets whereas Player 1 has $1+2\cdot 10=21$. For its second turn, Player 1 has ten options depending on which outcome of $e_2$ it observed: $\Pi_{\cC} = \{ \pi^i_{\cC} \}_{i=1}^{10}$ and $\Pi_{\cD} = \{ \pi^i_{\cD} \}_{i=1}^{10}$. See the full version of this paper for an illustration. 
Figure~\ref{fig:game2} provides the extenstive-form game tree for \otg

For each instance of \otg and each $\pi^i_{\cA}$ (respectively, $\pi^i_{\cB}$), we sampled $P(C \mid A, \pi^i_{\cA})$ (respectively, $P(C \mid B, \pi^i_{\cB})$) from $U[0,1]$. Each leaf utility was chosen uniformly at random from the set $\{0, 0.1, \dotsc, 9.9, 10\}$. We experimented with two game model forms: one for when the simulator returned a noisy payoff vector and $e_1$ only, and one for when it returned the vector and outcomes of both events.

The final game was \fourRound, which begins with a stochastic event $e_1 \in \{ A, B, C, D \}$. Player 1 observes the event and then takes a turn, choosing one of four possible actions. Next, Player 2 observes the event (but not Player 1's action) and also chooses from four possible actions. This 3-round sequence is repeated twice, but in each subsequent sequence, the only outcomes available to Nature and the agents are the remaining ones that have not yet been chosen.
For instance, if $e_1 = A$, then Nature can only output $e_2 \in \{ B, C, D \}$ during its second turn and $e_3 \in \{ B, C, D \} \setminus \{ e_2 \}$ during its third. Likewise, the players are restricted to the actions that they have not yet played in the previous 3-round sequence(s). Since the players are only unable to observe the other player's actions during the \textit{current} 3-round sequence, each player has $4 + 4^3 \cdot 3 + 4^3 \cdot 3^3 \cdot 2 = 3652$ information sets. To compare the effects of varying degrees of tree exploitation, we examined three different game model forms: (1) simulator reports observation $e_1$ only; (2) simulator reports $e_1$ and $e_2$ only; and (3) simulator reports all three events. 
We believe that a model that includes only the first stochastic event would generally yield only a negligible difference in accuracy from a model that includes only the second (or third) stochastic event.

Each iteration of EGTA had a fixed budget of 500 total samples available for all strategy combinations to be fed into the simulator for \tg and \otg. Due to the larger size, we allotted 5000 total samples for \fourRound. We ran the experiments for \tg on a standard laptop (Quad-Core Intel Core i7 Processor, 2.7 GHz, 16GB RAM). Each repetition of both TE-PSRO and NF-PSRO for \tg finished in less than 1 min. 
We ran the experiments for \otg and \fourRound on a single core of the Great Lakes Slurm cluster at the University of Michigan, with 786MB of memory.
NF-PSRO on \otg consistently finished within 6 minutes, and took 4--90 minutes for \fourRound. TE-PSRO  required between 3 minutes and 5 hours for \otg (depending on the MSS used, see \S\ref{sec:exp2}), and at most 1 hour for \fourRound. 
All figures include the metrics' initial values at time-step~0.

\subsection{TE-EGTA Payoff Estimation}\label{sec:exp1}
 
 \begin{figure*}[ht!]
    \centering
    \begin{subfigure}[b]{0.49\textwidth}
     \includegraphics[width=0.9\textwidth]{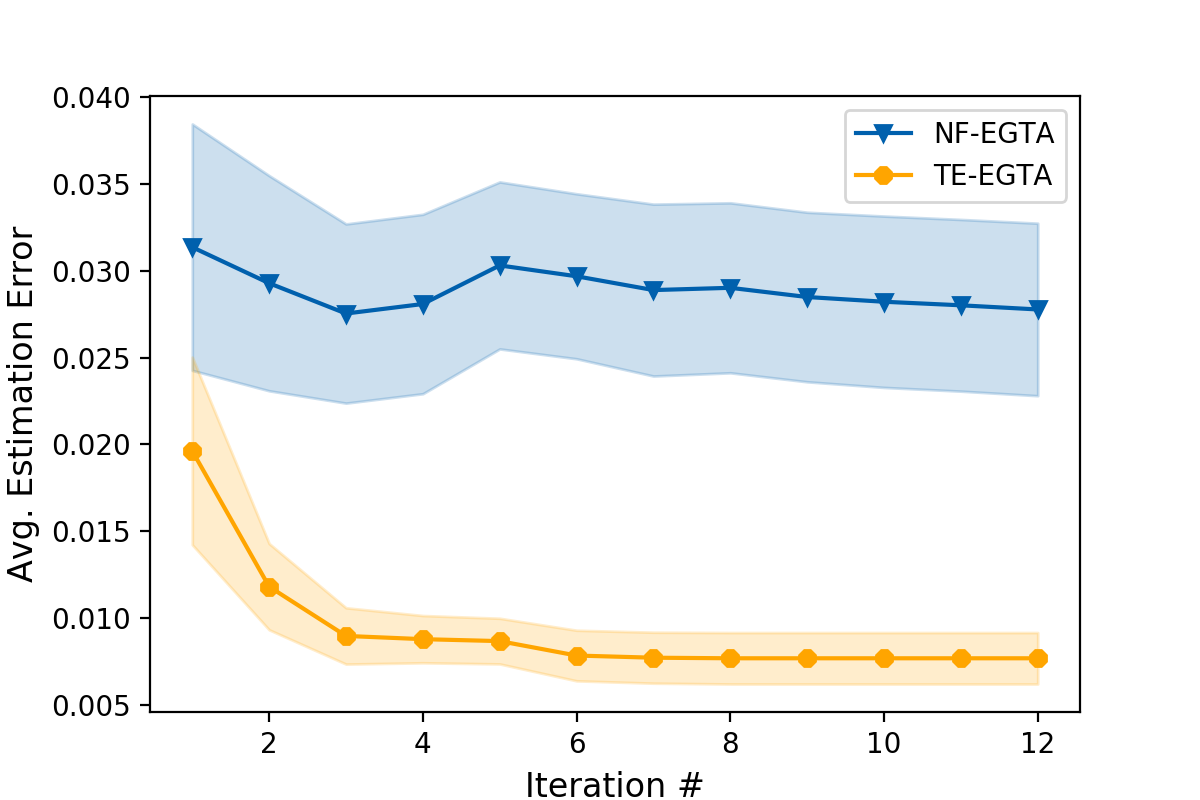}
    \caption{\tg \label{fig:game1_err}}
    \end{subfigure}~
    \begin{subfigure}[b]{0.49\textwidth}
     \includegraphics[width=0.9\textwidth]{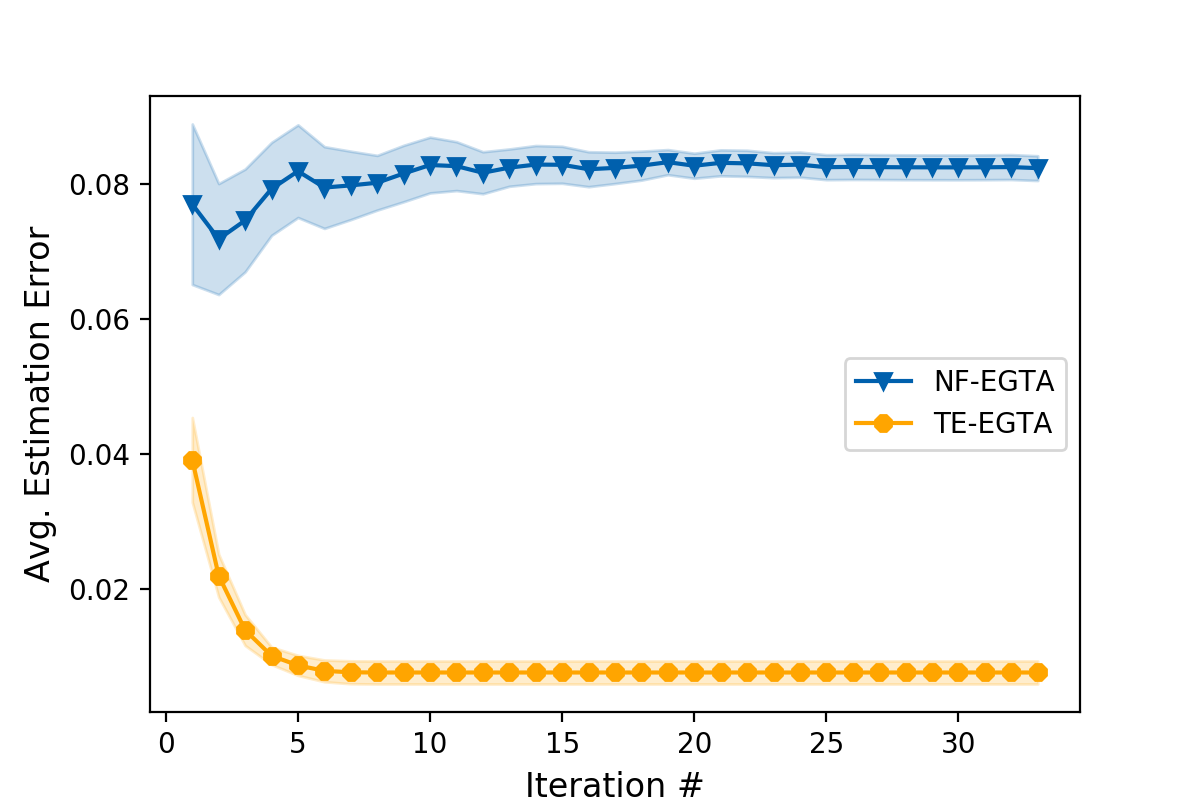}
    \caption{\otg \label{fig:game2_err}}
    \end{subfigure}\\
    \begin{subfigure}[b]{0.49\textwidth}
    \includegraphics[width=0.9\textwidth]{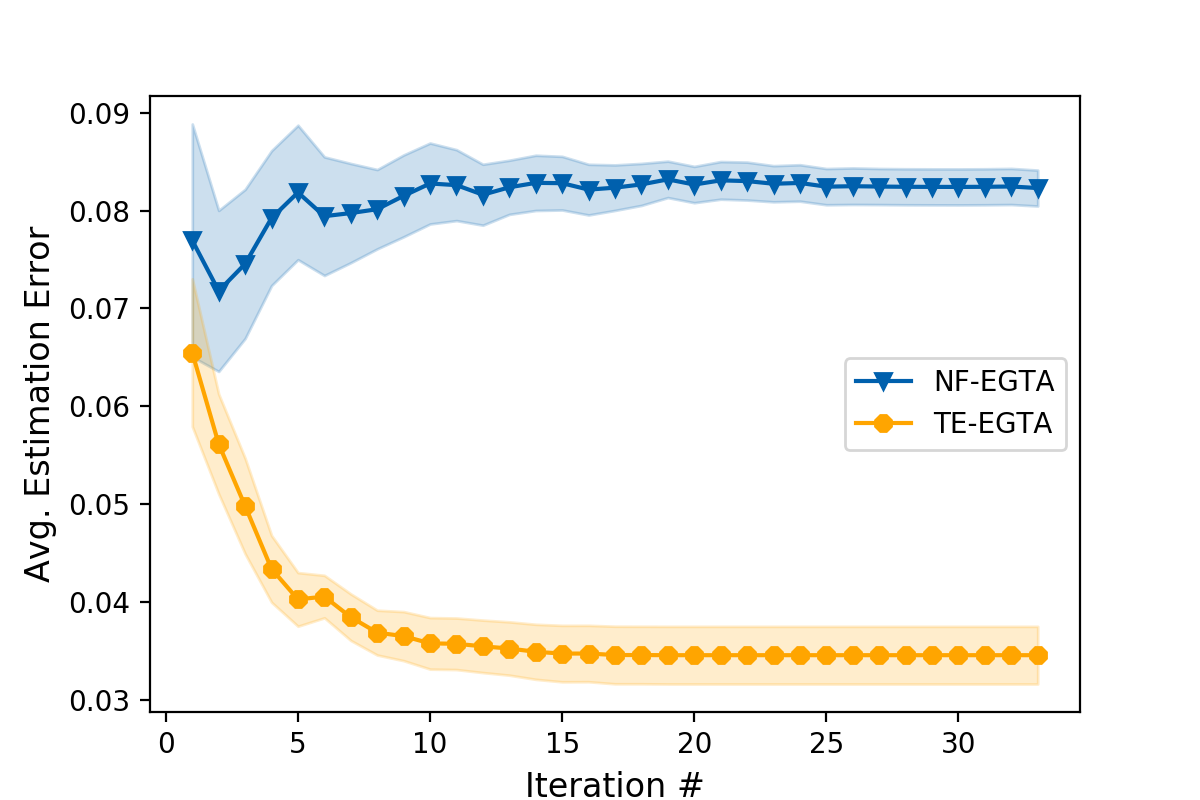}
    \caption{\otg with 1 event \label{fig:g2g1_err}}
    \end{subfigure}~
    \begin{subfigure}[b]{0.49\textwidth}
    \includegraphics[width=0.9\textwidth]{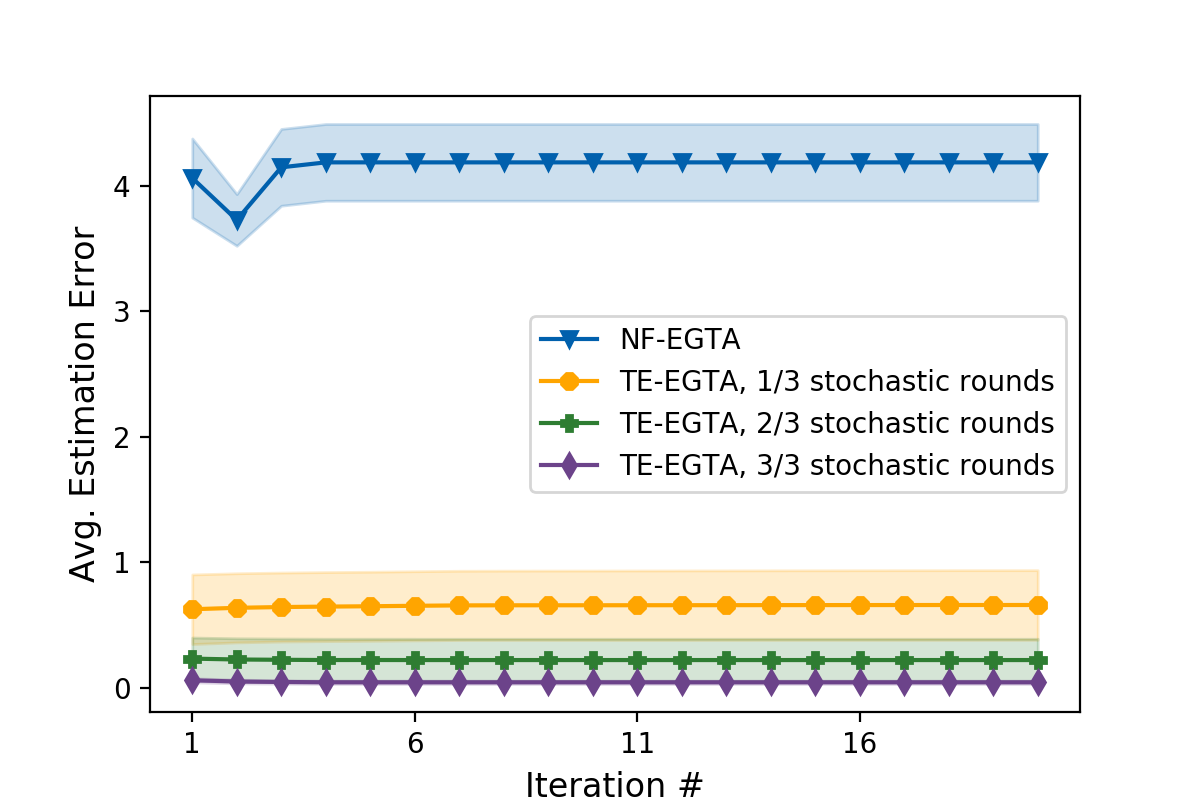}
    \caption{\fourRound \label{fig:game3_err}}
    \end{subfigure}~
    \caption{Average estimation error of strategy payoffs over the course of EGTA's runtime. \label{fig:est_err}
    Shaded areas represent the  standard error of the mean. The estimation errors at iteration 0 are identical since the restricted sets for both models contain the same randomly chosen policy; hence, they are omitted.}
\end{figure*}
 
 The aim of the first set of experiments was to assess the improvement in strategy profile payoff estimation produced by incorporating the EFG tree structure into the empirical game model. 
 We ran NF-EGTA and TE-EGTA on each true game with the same number $m=500$ of simulations for each strategy-profile payoff vector estimation. To update the game model for either variant of EGTA, we implemented the PSRO framework using an oracle that returns the best response to the other player's strategy for \tg and \otg. However, the size of \fourRound made a best response oracle infeasible, so we instead used Q-learning to compute an approximate best response from the true game. For newly selected strategy profiles that were simulated in each iteration, we computed estimated payoffs $\hat{U}^{\NF}_j(\bsigma)$ (resp. $\hat{U}^{\TE}_j(\bsigma)$) for NF-EGTA (resp. TE-EGTA) from accumulated simulation data using the approach described in \S\ref{sec:NF_EGTA} (resp.~\S\ref{sec:TE_EGTA_param_est}). 
 We evaluated the \textit{estimation error} for that iteration of either variant as the average absolute difference between true and estimated payoffs for all players over all strategy combinations in the current empirical game. We repeated this operation for 25 initial restricted sets, each consisting of a single randomly chosen policy, and reported the estimation error averaged over all 25 repetitions for each iteration of PSRO in Fig.~\ref{fig:est_err}.
 
 As the plots show, TE-EGTA achieves significantly lower payoff estimation error compared to NF-EGTA across all games.
 It is also clear that while the vast number of infosets in \fourRound led NF-EGTA to perform worse as more strategy combinations were added despite an unchanging sample budget~$m$; such was not the case for TE-EGTA, which converged very quickly. 
 We attribute this to the relatively small number of actions (2, 3, or 4) available at each information set, as well as the large number of infosets relative to the total number of game paths. Q-learning returned a best response for every infoset that could be reached, given~$\bm{\sigma}$, so the empirical game ceased growing after only a few iterations. 
 Finally, we note that the more stochastic events included in $\hat{G}$, the more tree structure is exploited by TE-EGTA, and the lower the resulting payoff error.
 In fact, the inclusion of even a single stochastic event or round in the model dramatically decreased the payoff error in comparison to NF-EGTA.

\subsection{Iterative Model Refinement in PSRO}\label{sec:exp2}
Our second set of experiments compared the power of NF-PSRO and TE-PSRO to iteratively explore the EFG's strategy space and fine-tune their respective empirical game models. 
PSRO terminates once no new best responses can be added to $\hat{\Sigma}$. To evaluate the efficacy of this iterative fine-tuning, we computed the regret $\regret(\bsigma)$ (as defined in \S\ref{sec:EFG_model}) in the true game $G$ of the solution $\bsigma$ returned by the MSS in every iteration. 

\begin{figure*}[ht!]
    \centering
    \begin{subfigure}[b]{0.49\textwidth}
     \includegraphics[width=0.9\textwidth]{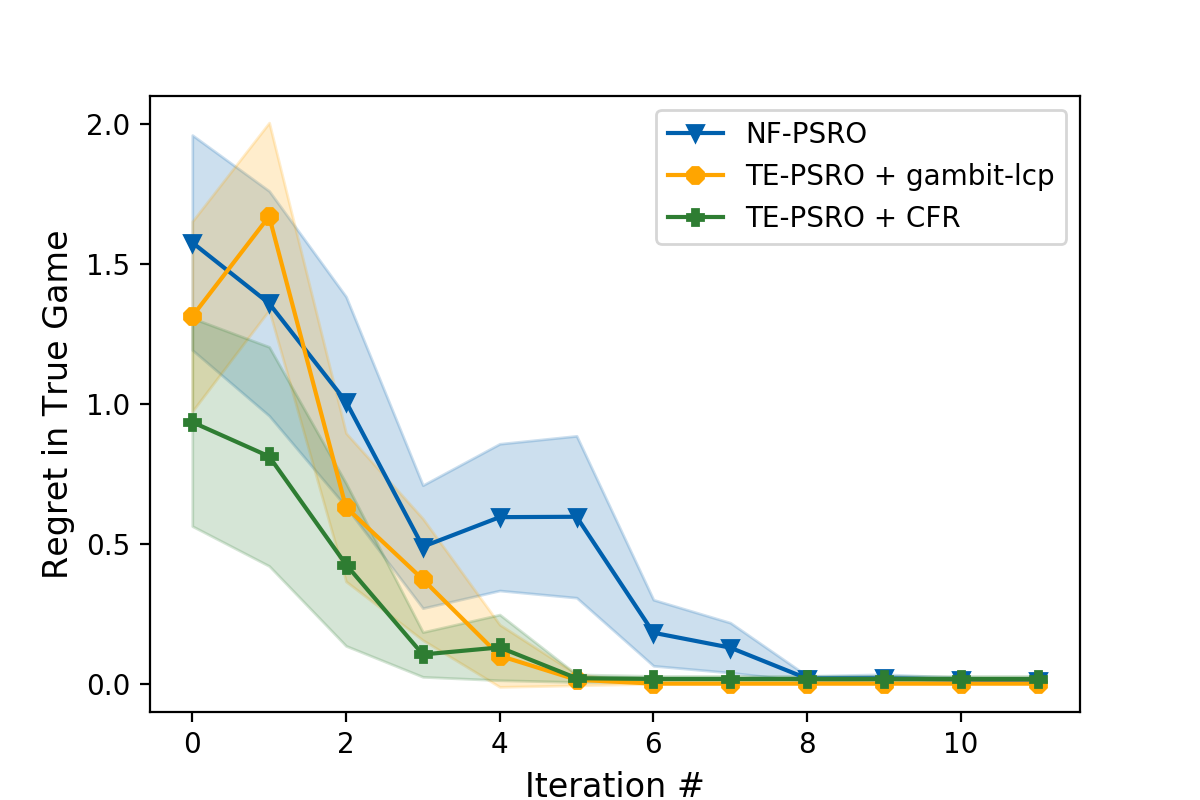}
    \caption{\tg \label{fig:game1_reg}}
    \end{subfigure}~
    \begin{subfigure}[b]{0.49\textwidth}
     \includegraphics[width=0.9\textwidth]{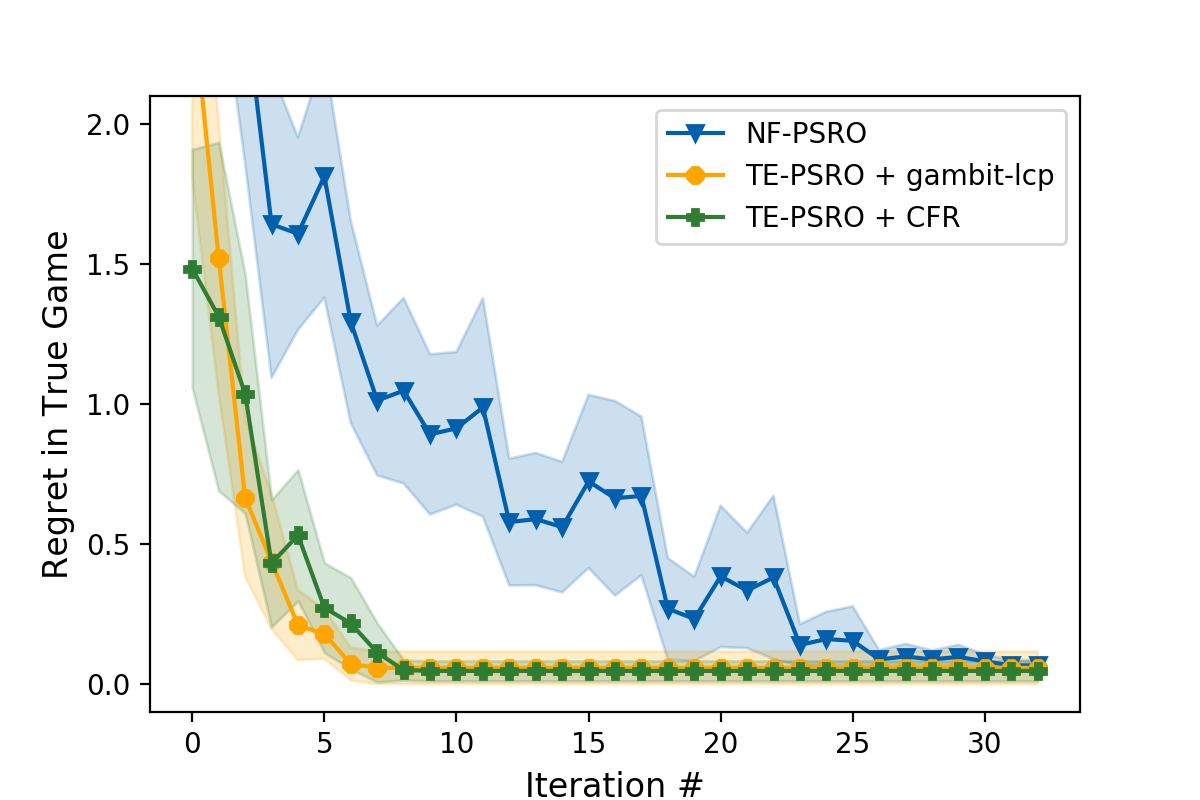}
    \caption{\otg \label{fig:game2_reg}}
    \end{subfigure}\\
    \begin{subfigure}[b]{0.49\textwidth}
     \includegraphics[width=0.9\textwidth]{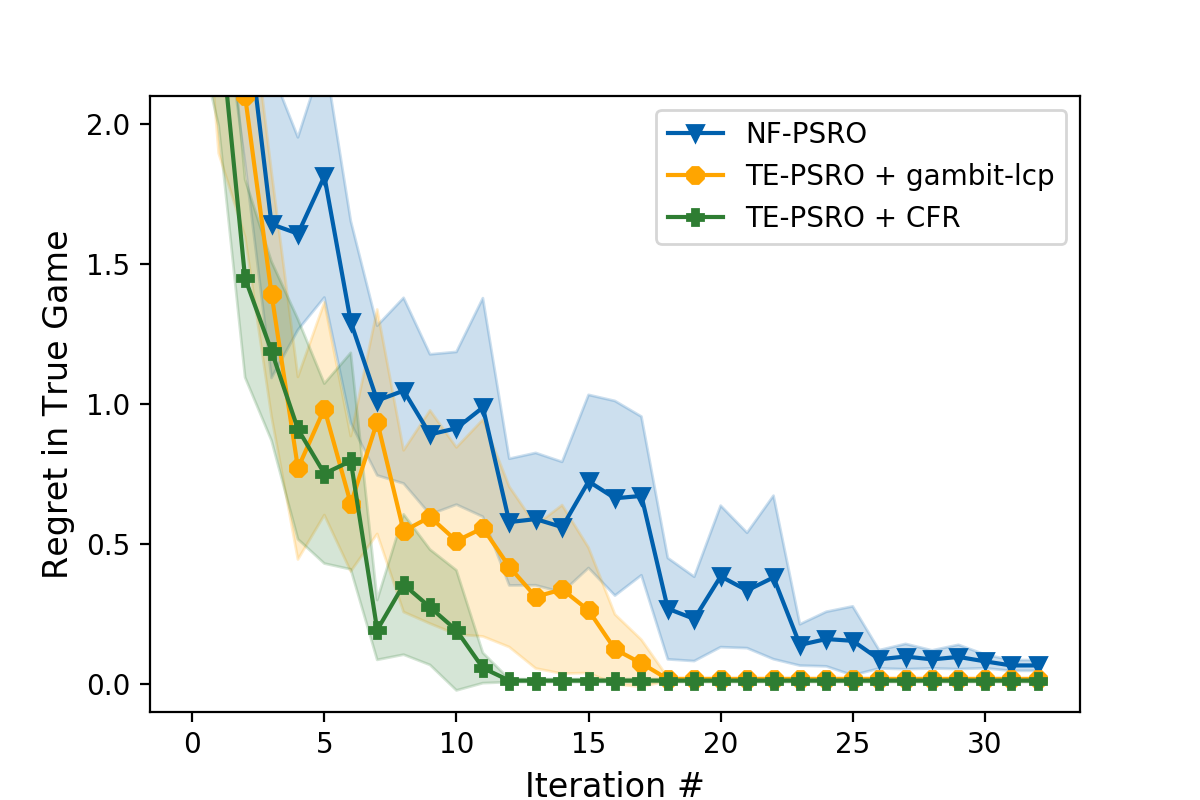}
    \caption{\otg with 1 event \label{fig:g2g1_reg}}
    \end{subfigure}~
    \begin{subfigure}[b]{0.49\textwidth}
     \includegraphics[width=0.9\textwidth]{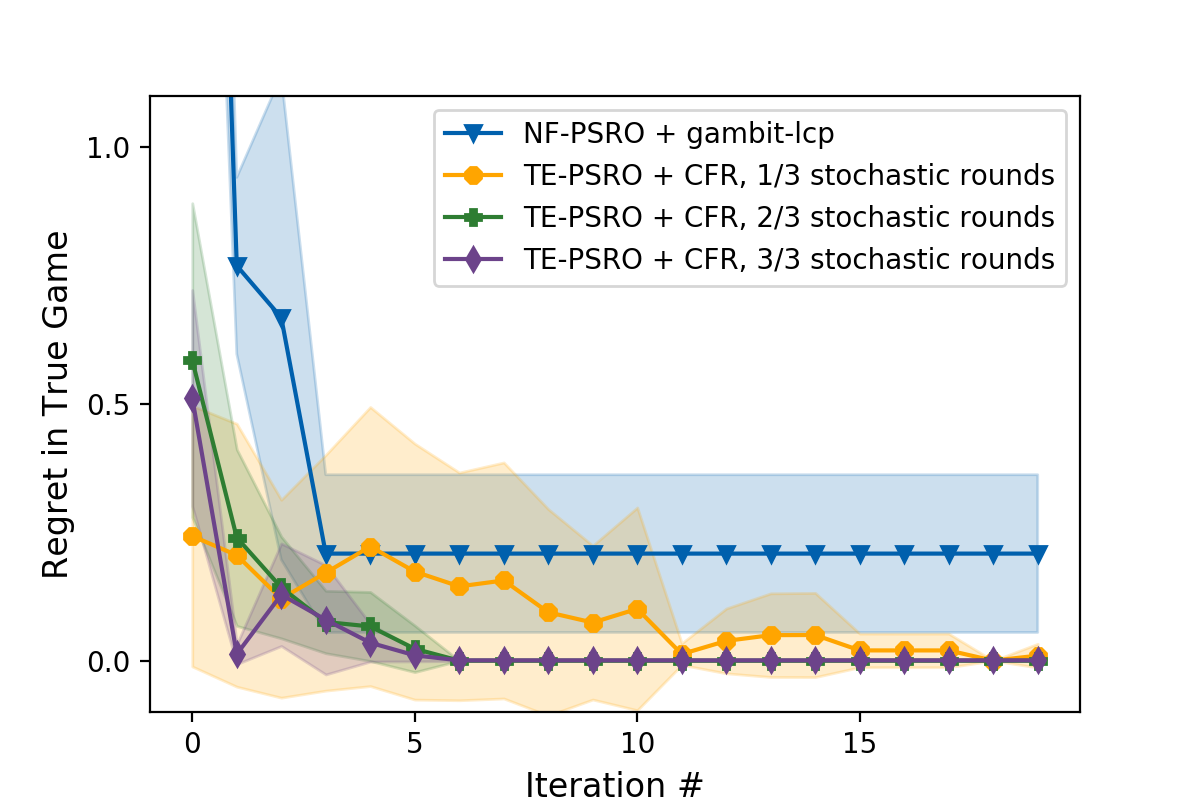}
    \caption{\fourRound \label{fig:game3_reg}}
    \end{subfigure}
    \caption{Average regret of solution profiles over the course of PSRO's runtime. 
    Shaded areas represent standard error of the mean. \label{fig:reg}}
\end{figure*}

For NF-PSRO, we used the Python-Gambit interface to represent the empirical game 
and used Gambit's \texttt{lcp} solver as the MSS. The solver takes as input an NFG or EFG, converts it into a linear complementarity program, and solves for all NE. We also used the \texttt{lcp} as the TE-PSRO solver for \tg and \otg. It is important to note that Gambit's solvers can become intractable for medium or large game trees.
However, when possible, we intentionally chose an MSS that finds exact solutions to the empirical game in order to minimize any error/variability in the solutions resulting from the iterative process of adding strategies and fine-tuning the empirical game models. For medium-to-large game trees like \fourRound, we used counterfactual regret minimization (CFR) \citep{cfrm07} to find an approximate NE and Q-learning to learn an approximate best response from the true game. We used CFR as the MSS for \otg as well for comparison to the exact \texttt{lcp} solver. 
As in~\S\ref{sec:exp1}, we repeated PSRO for 25 different restricted sets, each consisting of a single, randomly chosen strategy profile. We report the regret curves, averaged over 25 repetitions, in Fig.~\ref{fig:reg}.


TE-PSRO converged on average to a regret at least as tight as NF-PSRO using the same simulation budget and regardless of which pure $\bsigma$ the initial restricted set contained. 
It also converged in fewer iterations, particularly in \otg and \fourRound as more tree structure was included in $\hat{G}$. Additional plots in the full version of this paper demonstrate the same result for different numbers of samples. 
However, the standard error shadings for \tg 
overlap mainly due to the high volatility in NF-PSRO regret in earlier iterations. Since, in each iteration, we add new, pertinent best responses to $\hat{\Sigma}$, we hypothesize that their absence from the previous strategy space caused the regret to increase. A one-sided two-sample t-test on each of the iterations of \tg's regret curves established that TE-PSRO's regret improvement was statistically significant. 
These results suggest that including even some tree structure in $\hat{G}$ results in PSRO converging at least as quickly and to a solution that has lower regret in the true game. 

\section{Conclusions and Future Work}\label{sec:disc}
This study represents a first step towards the goal of leveraging extensive-form structure within the EGTA framework. 
 Our work complements prior research that showed benefits of exploiting tree structure in game reasoning and learning, for example studies that demonstrated advantages of extensive form in techniques based on the double oracle algorithm \citep{bosansky2014exact,McAleer21}.
 In future work, we hope to draw on further insights from this line of work, combining the best features of techniques from game reasoning, machine learning, and simulation-based game modeling. 
 One particularly fruitful direction may be consideration of strategy exploration methods that explicitly consider extensive structure in the currently defined strategy space.

\subsubsection{Acknowledgments} This work was supported in part by a grant from the Effective Altruism Foundation, and by the US National Science Foundation under CRII Award 2153184.
%
%
%
\bibliographystyle{abbrvnat}
\bibliography{tree_egta.bib}
%
\newpage
\appendix

\section*{Appendices (Supplemental)}

\section{Omitted proofs}\label{app:proofs}
\subsection{Proofs for Section~\ref{sec:NF_EGTA}}\label{app:NFG_unbiased}
Recall from Section~\ref{sec:NF_EGTA} that, to estimate a strategy-profile payoff vector $U(\bm{\sigma})$ in each NF-EGTA iteration, we simulate the strategy profile $m$ times, and hence compute the averages 
\[ \hat{U}^{NF}_j(\bm{\sigma}) = \frac{1}{m} \sum_{i = 1}^{m} \bar{u}^i_j \quad \forall j \in N \setminus \{0\}, \] 
where $\bar{u}^i_j$ denotes the realized payoff sample of each player $j \in N \setminus \{0\}$ at the end of the $i^\mathrm{th}$ simulation, $i = 1,2,\dots,m$.

\noindent \textbf{Restatement of Proposition~\ref{prop:unbiased_NF}.}
\emph{The NF-EGTA estimate $\hat{U}^{NF}_j(\bm{\sigma})$ is an unbiased estimator of the true strategy-profile payoff, i.e. $\mathbb{E}\left[ \hat{U}^{NF}_j(\bm{\sigma}) \right] = U_j(\bm{\sigma})$ for every player $j = 1,2,\dots,n$.}

\begin{proof}
Given any strategy profile $\bsigma$, 
suppose an arbitrary undisclosed $t \in T$ is reached in $m_t$ out of these $m$ simulated game-plays (each corresponding to a path).  We do not observe any $m_t$, but we do know that $\sum_{t \in T} m_t = m$ and that each terminal node $t \in T$ is reached with a probability $r(t, \bsigma)$ by the definition of $r(t, \bsigma)$ from Section~\ref{sec:NF_EGTA}. Thus, each $m_t$ is a priori binomially distributed with parameters $m$ equal to the number of trials and $r(t,\bsigma)$ equal to the per-trial fixed probability of reaching terminal node $t$ (i.e., a success).
Hence, $\expect\left[ m_t \right] = m \cdot r(t, \bsigma)$ for every $t \in T$.

Denote player $j$'s realized payoff sample at the end of the $i^\mathrm{th}$ of these $m_t$ simulations ending at node $t$ by $\bar{u}^i_j(t)$. By the property of the simulator and linearity of expectation, $\expect\left[ \bar{u}^i_j(t) \right]=u_j(t)$ for every $j \in N \setminus \{0\}$. 
It follows that we can rewrite the NF-EGTA payoff estimates as follows and compute the expected value with respect to the pertinent terminal nodes:
\begin{align*}
\hat{U}^{\NF}_j(\bsigma) &= \frac{1}{m} \sum_{t \in T} \sum_{i = 1}^{m_t} \bar{u}^i_j(t).\\
\text{Hence,} \quad \expect_{t \sim r(T, \bsigma)}\left[ \hat{U}^{\NF}_j(\bsigma) \right] &= \frac{1}{m} \sum_{t \in T} \expect\left[ \sum_{i = 1}^{m_t} \bar{u}^i_j(t) \right] \\
&= \frac{1}{m} \sum_{t \in T} \expect\left[ m_t \right] \cdot \expect\left[ \bar{u}^i_j(t) \right] \\
&= \frac{1}{m} \sum_{t \in T} (m \cdot r(t, \bsigma)) u_j(t)\\ 
&= \sum_{t \in T} u_j(t) \cdot r(t, \bsigma)\\
&= U_j(\bsigma).
\end{align*}
The first equality follows simply from the linearity of expectation; the second holds since the quantity of summands $\bar{u}^i_j(t)$ is a binomial random variable $m_t$; and the final two equalities follow from the definition of $U_j(\bsigma)$. \hfill $\square$
\end{proof}

\subsection{Proofs for Section~\ref{sec:TE_EGTA_param_est}}\label{app:TEEGTA_payoff_est}

Recall, from Section~\ref{sec:TE_EGTA_param_est}, the formula for computing the TE-EGTA payoff estimate for each player $j$'s under strategy profile $\bsigma$:
\begin{align*}
    \hat{U}^{\TE}_j(\bsigma) 
    &= \sum_{t \in T} \hat{u}_j(t) \prod_{k = 1}^n r_k(t, \sigma_k) \left( \prod_{w \in \varphi(t, 0)} \frac{m_w}{m_{\parent[w]}} \right).
\end{align*}
where $m_h$ is the number of times out of all $m$ simulations that a node $h$ is reached across all strategy profiles in the restricted set.

\noindent \textbf{Restatement of Proposition~\ref{prop:teegta_est}.} \emph{For every player $j \in N \setminus \{0\}$ and strategy profile $\bsigma \in \hat{\Sigma}$, $\expect_{t \sim r(T, \bsigma)} \left[ \hat{U}^{\TE}_j(\bsigma) \right] = U_j(\bsigma)$.}

\begin{proof}
Conditioned on a terminal node $t$, $\hat{u}_j(t)$ and $\prod_{w \in \varphi(t, 0)} \frac{m_w}{m_{\parent[w]}}$ are independent random variables. The randomness in the first is due to uncertainty in the simulator's utility output, given by a symmetric distribution (such as Gaussian) centered around the true leaf utility. The randomness in the second is due to the uncertainty in the path traversed during an actual instantiation of the strategy profile (including the stochasticity in Nature's choice).

Hence, by the sum law of expectations, $\expect \left[ \hat{u}_j(t) \right] = u_j (t)$. We note also that
\begin{equation*}
\prod_{w \in \varphi(t, 0)} \frac{m_w}{m_{\parent[w]}} = \prod_{i = 1}^{\abs{\varphi(t, 0)}} \frac{m_{w_i}}{m_{w_{i-1}}}
\end{equation*}
where $m_{w_0} \equiv m_{\varphi(t)[0]}$ is the total number of times out of $m$ that the first node in the path $\varphi(t)$ is reached, and $w_1, \dotsc, w_{\abs{\varphi(t, 0)}}$ is the list of chance nodes along the path from the root to $t$. We note that this expression can be rewritten as
\begin{equation*}
    \hat{r}_0(t) = \prod_{i = 1}^{\abs{\varphi(t, 0)}} \frac{m_{w_i}}{m_{w_{i-1}}} = \frac{m_{w_{\abs{\varphi(t, 0)}}}}{m_{\varphi(t)[0]}}
\end{equation*}
where $m_{\abs{\varphi(t, 0)}}$ is the final node in the path and $m_{w_{\abs{\varphi(t, 0)}}} \sim \Binom(m, r_0(t, \bsigma))$.

Finally, taking the expectation of $\hat{U}^{\TE}_j(\bsigma)$ with respect to all the above sources of uncertainty and applying the linearity property of expectation for independent random variables,
\begin{align*}
    \expect \left[\hat{U}^{\TE}_j(\bsigma) \right] &= \sum_{t \in T} \expect \left[ \hat{u}_j(t) \cdot \prod_{k = 1}^n r_k(t, \sigma_k) \cdot \prod_{w \in \varphi(t, 0)} \frac{m_w}{m_{\parent[w]}} \right]\\
    &= \sum_{t \in T} \expect \left[ \hat{u}_j(t) \right] \cdot \prod_{k = 1}^n r_k(t, \sigma_k) \cdot \expect  \left[ \prod_{w \in \varphi(t, 0)} \frac{m_w}{m_{\parent[w]}} \right]\\
    &= \sum_{t \in T} \expect \left[ \hat{u}_j(t) \right] \cdot \prod_{k = 1}^n r_k(t, \sigma_k) \cdot \expect  \left[ \frac{m_{w_{\abs{\varphi(t, 0)}}}}{m_{\varphi(t)[0]}} \right]\\
    &= \sum_{t \in T} u_j(t) \cdot \prod_{k = 1}^n r_k(t, \sigma_k) \cdot \frac{m_{\varphi(t)[0]} \cdot r_0(t, \bsigma)}{m_{\varphi(t)[0]}}\\
    &= \sum_{t \in T} u_j(t) \cdot \prod_{k = 1}^n r_k(t, \sigma_k) \cdot r_0(t, \bsigma)\\
    &= \sum_{t \in T} u_j(t) \cdot r(t, \bsigma) = U_j(\bsigma).
\end{align*}
The last equality follows from Equation~\eqref{eq:te_util_exp} in Section~\ref{sec:tree_expl_egta}. \hfill $\square$
\end{proof}

\subsection{Proofs for Section~\ref{sec:theo_improv}}\label{app:main_results}
\noindent \textbf{Restatement of Proposition~\ref{prop:uni_approx_reg}.}
\emph{If $\hat{G}$ is a uniform $\varepsilon$-approximation of $G$ and $\bsigma$ is a $\gamma$-Nash equilibrium of $\hat{G}$ for some $\gamma \ge 0$, then $\regret_j(\bsigma) \le 2\varepsilon + \gamma$ for each player $j \in N \setminus \{0\}$ upon the termination of EGTA.}

\begin{proof}
We adapt the proof of \citet[Theorem 2.2]{viqueria19} to our setting.

For the strategy profile $\bsigma$ under consideration, let
\begin{align*}
    \bsigma^{*} \in \argmax_{\sigma_j \in \hat{\Sigma}_j}~U_j(\sigma_j, \bsigma_{-j}); \qquad
    \hat{\bsigma}^{*} \in \argmax_{\sigma_j \in \hat{\Sigma}_j}~\hat{U}_j(\sigma_j, \bsigma_{-j}).
\end{align*}

Recall that any strategy $\sigma_j$ induces a probability distribution over $\Pi_j(I)$ for each information set $I$ of player $j$. Recall also that $\hat{\Sigma}_j \subseteq \Sigma_j$ for any player $j$. We wish to demonstrate that
\begin{align*}
    U_j(\bsigma^{*}) = \max_{\sigma_j \in \hat{\Sigma}_j}~U_j(\sigma_j, \bsigma_{-j}) = \max_{\sigma_j \in {\Sigma}_j}~U_j(\sigma_j, \bsigma_{-j}).
\end{align*}
In order to do this, for each player $j$, it must be true that the policy $\sigma'_j$ that maximizes $j$'s utility is included in the empirical restricted set $\hat{\Sigma}_j$ by the time that EGTA terminates. Since we add new policies to the restricted set $\hat{\Sigma}_j$ for each player using best response, the policy in question falls into one of three possible cases:
\begin{enumerate}
    \item Policy $\sigma'_j \in \hat{\Sigma}_j$ and is part of the support of the final optimal solution $\bsigma^{*}$;
    \item Policy $\sigma'_j \in \hat{\Sigma}_j$ but is not part of $\bsigma^{*}$'s support;
    \item Policy $\sigma'_j \notin \hat{\Sigma}_j$.
\end{enumerate}

Case 1 is trivial. Case 2 means that there exists a better mixed strategy for player $j$ such that there is no incentive to deviate to another strategy such as $\sigma'_j$ in the restricted set, so $U_j(\hat{\bsigma}^{*}) \geq U_j(\sigma'_j, \hat{\bsigma}^{*}_{-j})$. We demonstrate that Case 3 is not a possible outcome at PSRO's termination through proof by contradiction. Assume that there exists $\sigma'_j \notin \hat{\Sigma}_j$, and that $\sigma'_j$ produces a utility greater than that of $\hat{\bsigma}^{*}$. If this is true, then $\sigma'_j$ must be the best response to $\hat{\bsigma}^{*}_{-j}$. It follows that $\sigma'_j$ must be added to the restricted set and PSRO must continue until neither player has new best responses that are not already in their respective restricted sets. Therefore, it is not possible for Case 3 to happen when PSRO has terminated, which means that $U_j(\bsigma^{*}) = \max_{\sigma_j \in \hat{\Sigma}_j}~U_j(\sigma_j, \bsigma_{-j}) = \max_{\sigma_j \in {\Sigma}_j} U_j(\sigma_j, \bsigma_{-j})$.

From the above equalities and the definition of player regret from Section~\ref{sec:EFG_model}, we get
\begin{align*}
    \regret_j(\bsigma) &= U_j(\bsigma^{*}) - U_j(\bsigma) \\
    &\leq \hat{U}_j(\bsigma^{*}) + \varepsilon - \left( \hat{U}_j(\bsigma) - \varepsilon \right) \\
    &\leq \hat{U}_j(\bm{\hat{\sigma}}^{*}) + \varepsilon - \left( \hat{U}_j(\bsigma) - \varepsilon \right) \\
    &\leq \hat{U}_j(\bm{\hat{\sigma}}^{*}) + \varepsilon - \left( \hat{U}_j(\bm{\hat{\sigma}}^{*}) - \varepsilon - \gamma \right) \\
    &\leq 2\varepsilon + \gamma.
\end{align*}
The first inequality follows from the fact that $\hat{G}$ is a uniform $\varepsilon$-approximation of $G$ (Section~\ref{sec:theo_improv} Definition~\ref{def:uniform_approx}), the second from the optimality of $\bm{\hat{\sigma}}^{*}$ as defined above, and the final line from the fact that $\bsigma$ is a $\gamma$-Nash equilibrium in $\hat{G}$. \hfill $\square$
\end{proof}

\noindent \textbf{Restatement of Theorem~\ref{thm:TE_tighter}.}
\emph{Under the assumptions of Section~\ref{sec:theo_improv}, for any $\delta \in (0,1)$ and the same number $m$ of game simulation repetitions in each iteration of either type of EGTA, there exist positive constants $\varepsilon_{\NF}$ and $\varepsilon_{\TE}$ such that 
$\frac{\varepsilon_{\TE}}{\varepsilon_{\NF}} = \frac{1}{\sqrt{c}}$,
and with probability at least $1 - \delta$ w.r.t. the randomness in the simulator payoff output, $\hat{G}_{\NF}$ (respectively, $\hat{G}_{\TE}$) is a uniform $\varepsilon_{\NF}$-approximation (respectively, $\varepsilon_{\TE}$-approximation) of $G$.}
\begin{proof}
Recall from Section~\ref{sec:EFG_model} 
that we assume $\hat{G}_{\NF}$ and $\hat{G}_{\TE}$ have the same resticted set $\hat{\Sigma}$. Without loss of generality, we assume that the noise for each sampled utility $\bar{u}_j(t)$ follows a sub-Gaussian distribution with a variance proxy of $\sigma^2_j$. We need to prove that TE-EGTA's empirical game model leads to a tighter $\ell_{\infty}$ norm than the normal-form model. 

First, we rewrite the payoff estimate computed by NF-EGTA as a sum over $m$ simulation iterations and all terminal nodes $t \in T$ using the Kronecker delta notation (where $t_i$ denotes the terminal node reached in the the $i^\mathrm{th}$ simulated play):

\begin{align*}
\hat{U}^{\NF}_j(\bsigma) &= \frac{1}{m} \sum_{t \in T} m_t \hat{u}_j(t) = \frac{1}{m} \sum_{i = 1}^m \sum_{t \in T} \bar{u}_j(t_i) \delta_{t_i t}.
\end{align*}

Next, recall the payoff estimate computed from the tree-exploiting model introduced in Section \ref{sec:theo_improv}, which relies on the direct estimation of parameters that are known to comprise the payoffs of different strategies and therefore are computed using the simulation data generated across several strategies in the restricted set. Each edge $(\parent[x], x)$ in the path from the root to some $t \in T$ is traversed with reach probability $r_{V(x)}(x, \sigma_{V(x)})$ or $r_k(t, \sigma_k)$ for $k \in N$. All of the reach probabilities $r_k(t, \sigma_k)$ for $k \neq 0$ are deterministic according to mixed strategy $\bsigma$, with the exception of any actions that are hidden in this example from player $j$ and instead are signaled through the stochastic observations from Nature whose reach probabilities $r_0$ must be estimated. The empirical probability of each edge in $\varphi(t)$ being traversed can also be expressed as a product of binomial ratios $\frac{m_w}{m_h}$ for $w \in \varphi(t, 0)$ and $h = \parent[w]$. 
The fractions cancel each other out, leaving $m'_t$ (the number of times terminal $t$ is reached out of $m$ samples in an iteration of TE-EGTA) in the numerator and some $cm$ in the denominator. $c$ is the number of strategy profiles from the restricted set that, after each is sampled $m$ times, result in a path taken through the tree that includes the first edge of $\varphi(t)$. 
$c$ is $O(1)$ for most games, but can expand to be as large as $O(\abs{\Sigma_j})$ for some $j \in N$ depending on the game structure and when the selected EGTA method terminates, depending on the size and format of the EFG\@. 
We combine this notion with the Kronecker delta to rewrite the estimated strategy payoffs for TE-EGTA:
\begin{align*}
    \hat{U}^{\TE}_j(\bsigma) &= \sum_{t \in T} \hat{u}_j(t) \cdot \prod_{k = 1}^n \hat{r}_k(t, \sigma_k) \prod_{w \in \varphi(t, 0)} \frac{m_w}{m_h} \\
    &= \sum_{t \in T} \hat{u}_j(t) \cdot \prod_{w \in \varphi(t)} \frac{m_w}{m_h} \\
    &= \sum_{t \in T} \hat{u}_j(t) \cdot \frac{m'_t}{c \cdot m} \\
    &= \sum_{t \in T} \frac{1}{c \cdot m} \left( \sum_{i = 1}^{m'_t} \bar{u}_j(t) \right) \\
    &= \frac{1}{c \cdot m} \sum_{i = 1}^{m} \sum_{t \in T} \bar{u}_j(t_i) \delta_{t_i t}.
\end{align*}

Using these expressions, we apply Hoeffding's inequality to give an upper bound for the probability that the empirical strategy payoffs differ from their expectations by a certain amount. Due to the presence of the Kronecker delta, the $i$-th term in each sum also falls within this bound. Additionally, because the noise function associated with each leaf utility $u_j(t)$ is sub-Gaussian with variance proxy $\sigma^2_j$, the $i$-th term of the summation $\sum_{t \in T} \bar{u}_j(t_i) \delta_{t_i t}$ for each expression is also sub-Gaussian with mean $u_j(t)$ and variance proxy $\sigma^2_j$. Note that this quantity is distinct and different from a complete strategy profile $\bm{\sigma}$, despite the similar notation.
The following bound therefore holds for all $j \in N$ when the normal-form game model is used to estimate the strategy payoffs:
\begin{equation*}
    \Prob\left( \abs{\hat{U}^{\NF}_j(\bsigma) - U_j(\bsigma)} \geq \varepsilon \right) \leq 2 \exp{\left(- \frac{2m^2 \varepsilon^2}{\sum_{i = 1}^m 2 \sigma^2_j} \right) } = 2 \exp{\left(- \frac{m \varepsilon^2}{2 \sigma^2_j} \right) }
\end{equation*}

Then with probability at least $1 - \delta$, the deviation $\varepsilon_{\NF}[j, \bsigma]:= \mid \hat{U}^{\NF}_j(\bsigma) - U_j(\bsigma) \mid$ between $\hat{U}^{\NF}_j(\bsigma)$ and $U_j(\bsigma)$ for $j \in N \setminus \{ 0 \} $ is bounded from above:
\begin{align*}
   \varepsilon_{\NF}[j, \bsigma] &\leq \sqrt{\frac{2 \sigma^2_j \log{\frac{2}{\delta}}}{m}}.
\end{align*}

Following the same approach, Hoeffding's inequality yields the following bound for all $j \in N$ when the tree-exploiting game model of TE-EGTA is used to estimate the strategy payoffs:
\begin{equation*}
    \Prob\left(\abs{\hat{U}^{\TE}_j(\bsigma) - U_j(\bsigma)} \geq \varepsilon \right) \leq 2 \exp{\left(- \frac{2c^2 m^2 \varepsilon^2}{\sum_{i = 1}^{cm} 2\sigma^2_j} \right) } = 2 \exp{\left(- \frac{c m \varepsilon^2}{2 \sigma^2_j} \right) }
\end{equation*}

With probability at least $1 - \delta$, the deviation $\varepsilon_{\TE}[j, \bsigma]:= \abs{\hat{U}^{\TE}_j(\bsigma) - U_j(\bsigma)}$ for $j \in N, j \neq 0$ is bounded from above:
\begin{equation*}
    \varepsilon_{\TE}[j, \bsigma] \leq \sqrt{\frac{2 \sigma^2_j \log{\frac{2}{\delta}}}{cm}}
\end{equation*}

Now when all players $j \in N$ and all strategies in the restricted set $\bsigma \in \hat{\Sigma}$ are considered, the following bounds result (note that the restricted set is assumed to be the same for both processes): 

\begin{align*}
    \max_{j \in N, \bsigma \in \hat{\Sigma}} \abs{\hat{U}^{\NF}_j(\bsigma) - U_j(\bsigma)} &\leq \sqrt{\frac{2 \sigma^2_j \log{\frac{2 \mid N \times \hat{\Sigma} \mid}{\delta}}}{m}} \equiv \varepsilon_{\NF} \\
    \max_{j \in N, \bsigma \in \hat{\Sigma}} \abs{\hat{U}^{\TE}_j(\bsigma) - U_j(\bsigma)} &\leq \sqrt{\frac{2  \sigma^2_j \log{\frac{2 \mid N \times \hat{\Sigma} \mid}{\delta}}}{cm}} \equiv \varepsilon_{\TE} \\
    \implies \frac{\varepsilon_{\TE}}{\varepsilon_{\NF}} &= \sqrt{\frac{1}{c}}.
\end{align*}
This completes the proof. \hfill $\square$
\end{proof}


\section{Detailed illustration of TE-PSRO (Section~\ref{sec:treePSRO})}\label{app:te_psro}
\paragraph{Expansion of the Empirical Game through TE-PSRO (\S\ref{sec:treePSRO})}

\begin{figure}[ht!]
    \centering
    \includegraphics[width=0.5\textwidth]{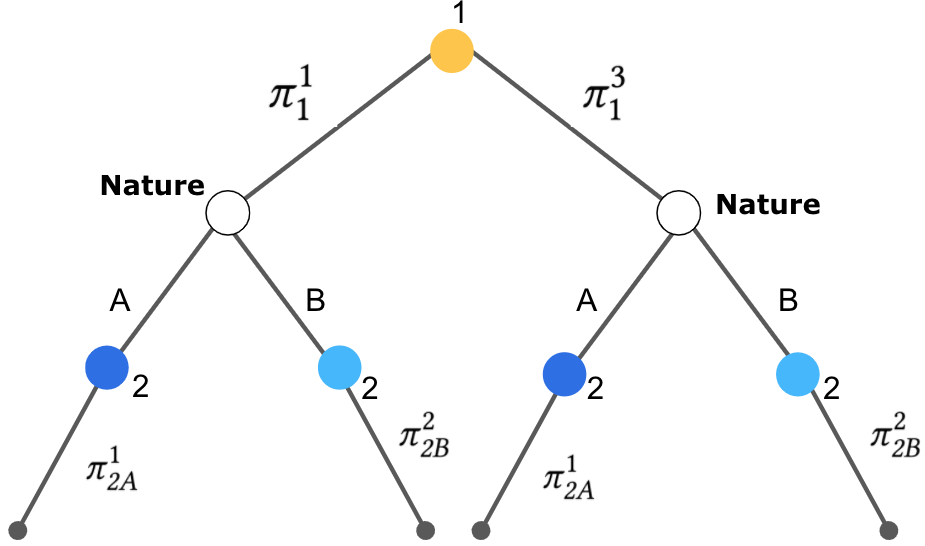}
    \caption{Sample 2-player empirical game with $\Pi_1 = \{ \pi^1_1, \pi^3_1 \}, \Pi_{\cA} = \{ \pi^1_{\cA} \}$, and $\Pi_{\cB} = \{ \pi^2_{\cB} \}$}
    \label{fig:te_psro_game}
\end{figure}

\begin{figure}[ht!]
    \centering
    \includegraphics[width=0.65\textwidth]{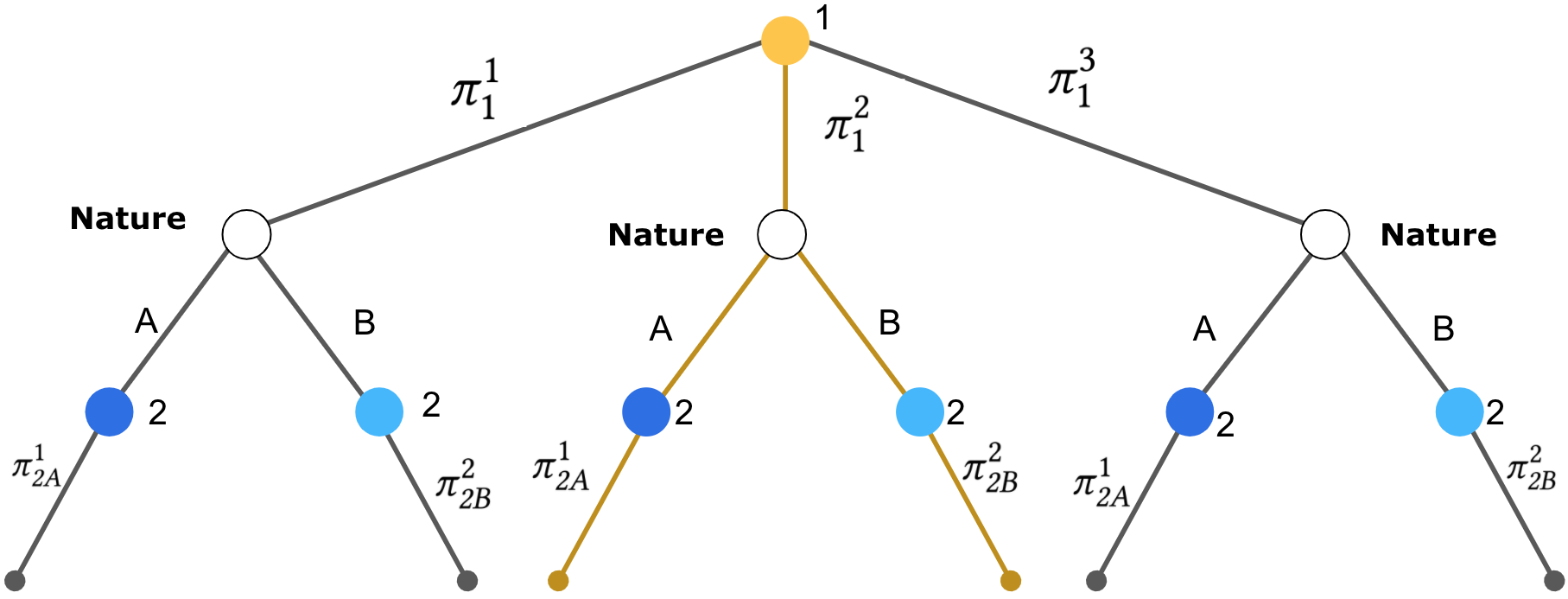}
    \caption{Empirical game model updated after simulating $\{ BR_1 \} \times \Pi_{\cA} \times \Pi_{\cB}$. The paths taken are given in gold.}
    \label{fig:first_mod}
\end{figure}

\begin{figure}[ht!]
    \centering
    \includegraphics[width=0.65\textwidth]{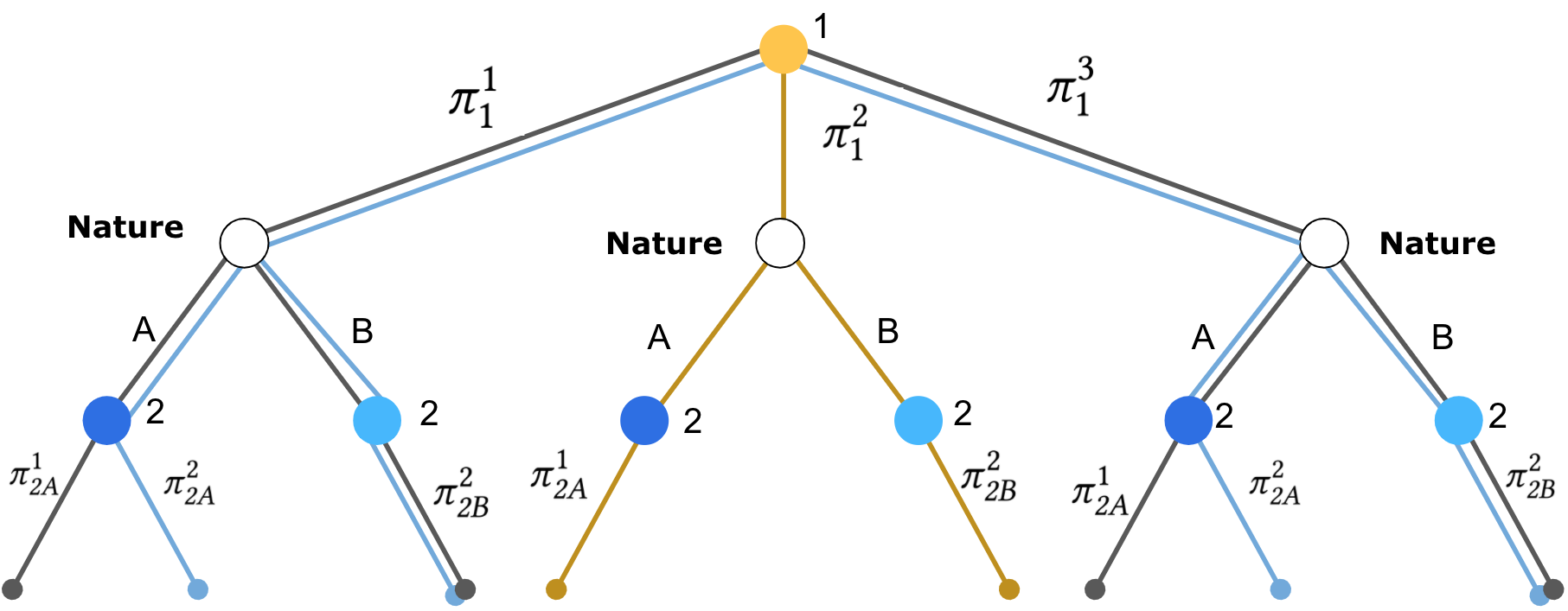}
    \caption{Empirical game model updated after simulating $\Pi_1 \times \{ BR_{\cA} \} \times \Pi_{\cB}$. The paths taken are given in blue.}
    \label{fig:second_mod}
\end{figure}

\begin{figure}[ht!]
    \centering
    \includegraphics[width=0.65\textwidth]{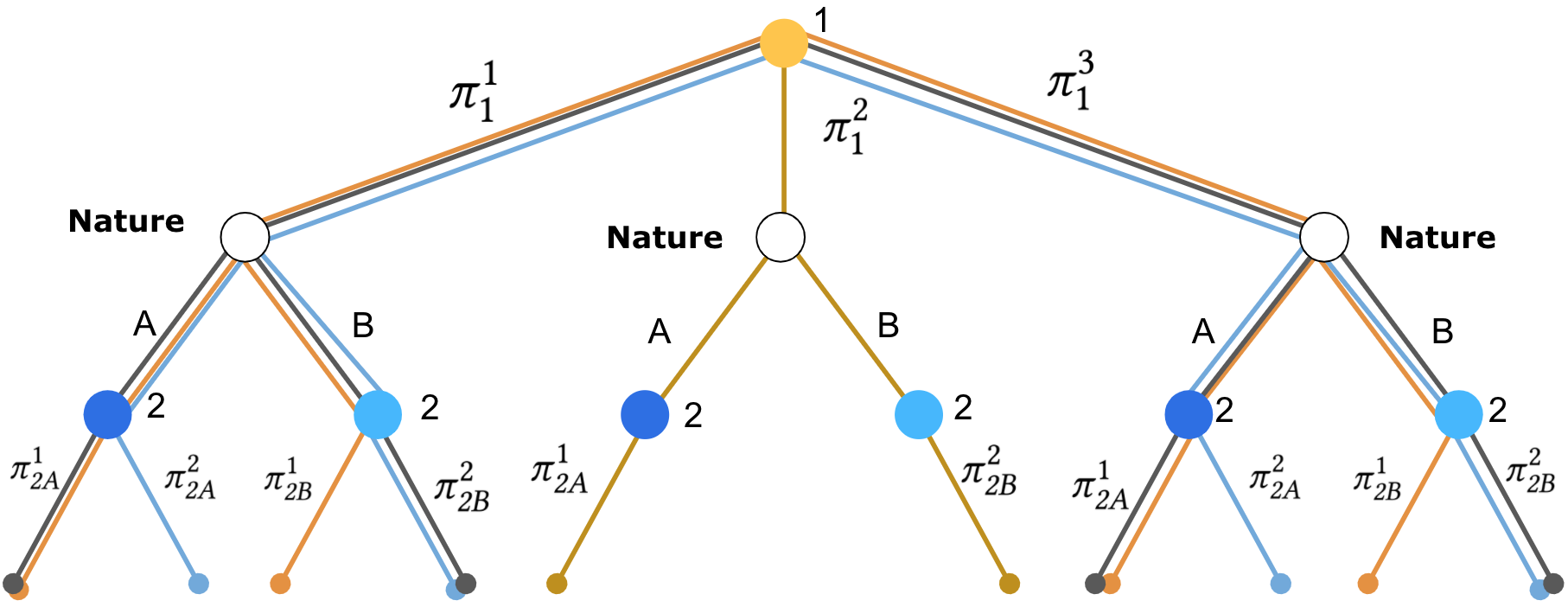}
    \caption{Empirical game model updated after simulating $\Pi_1 \times \Pi_{\cA} \times \{ BR_{\cB} \}$. The paths taken are given in orange.}
    \label{fig:third_mod}
\end{figure}

\begin{figure}[ht!]
    \centering
    \includegraphics[width=0.65\textwidth]{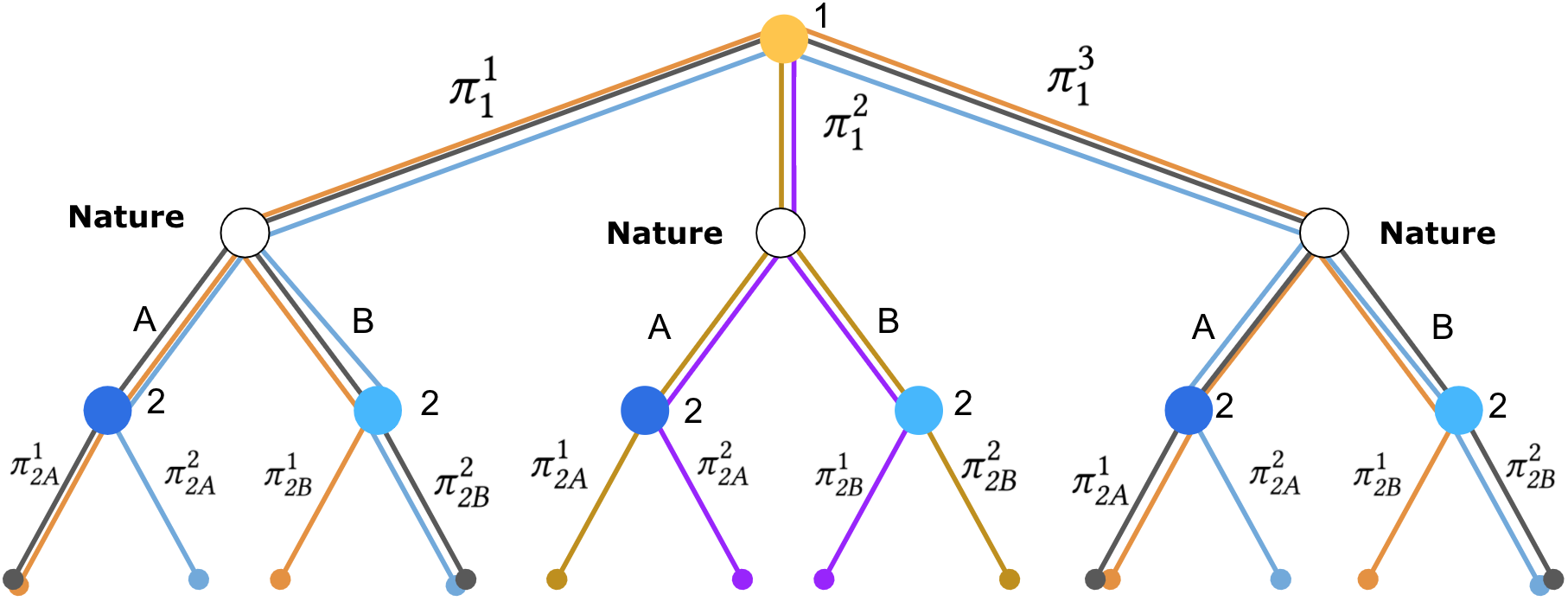}
    \caption{Empirical game model updated after simulating $(BR_1, BR_{\cA}, BR_{\cB})$. The paths taken are given in purple.}
    \label{fig:fourth_mod}
\end{figure}

We now illustrate how the best responses returned by PSRO are incorporated into the tree-exploiting empirical game model. Consider a simple empirical game illustrated by Fig. \ref{fig:te_psro_game} consisting of two strategies for player 1, one strategy for player 2 when event $A$ is observed, and one strategy for player 2 when event $B$ is observed. Suppose that the oracle returns $\pi^2_1$ as the best response for player 1 and $(\pi^2_{\cA}, \pi^1_{\cB})$ as the best response for player 2. Let $BR_1(\sigma_{\cA}, \sigma_{\cB})$ and $(BR_{\cA}(\sigma_1), BR_{\cB}(\sigma_1))$ denote these respective best responses to the current meta-strategy $(\sigma_1, (\sigma_{\cA}, \sigma_{\cB}))$. In NF-PSRO, each new strategy combination would result in a single new entry in the empirical payoff matrix. Figures \ref{fig:first_mod}-\ref{fig:fourth_mod} demonstrate how TE-PSRO expands the empirical game model as a result of simulating the new strategy combinations and organizing the simulation data to take advantage of the tree structure.

In Fig. \ref{fig:first_mod}, two new paths from root to leaf added in yellow as a result of simulating $BR^1$ with player 2's original strategy. In Fig. \ref{fig:second_mod}, $BR_{\cA}$ is simulated with player 1's original strategy and player 2's strategy when $B$ is observed. In Fig. \ref{fig:third_mod}, $BR_{\cB}$ is simulated with player 1's original strategy and player 2's strategy when $A$ is observed. Some paths are retread while some additional leaf nodes are added. In the case where paths are retread, more samples can be utilized to improve current estimates of old leaf utilities (such as $\mathbf{U}(\pi^1_1, B, \pi^2_{\cB})$) and conditional probabilities such as $\hat{P}(A \mid \pi^1_1)$. In Fig. \ref{fig:fourth_mod}, all best responses are simulated together in paths that partially overlap with those in Fig. \ref{fig:first_mod}. One can see that in a single step, the information extracted from the simulation data when the empirical game model gets updated is more complex and nuanced. It is also clear to see that like before, future best responses may overlap with the paths in this current game tree because the parameters are outlined by the tree itself, not the rows and columns of a payoff matrix as in NF-PSRO.

\newpage
\section{Omitted details and results from Section~\ref{sec:expts} (Experiments)}\label{app:expts}
\subsection{Note on MSS choice for Experiments in Section~\ref{sec:exp2}}\label{app:exp2_mss}
Initially, we attempted to use NashPy's Lemke-Howson algorithm as our MSS; however, we noticed that Lemke-Howson sometimes struggled to solve the empirical game in the intermediate iterations of PSRO, regardless of which empirical game model was used to compute the strategy payoffs. Sometimes, the experiments would halt because the empirical game was \textit{degenerate}, meaning there is an infinite number of mixed strategies for one player that are all the best response to the other player's strategy. This was not unexpected, as in the case of Kuhn poker, there are infinitely many mixed-strategy equilibria for the first player, who has to check or bet depending on what card he was dealt. Our choice of Gambit's \textsf{lcp} solver as the MSS avoids this problem.

\newpage
\subsection{Omitted plots}\label{app:supp_figs}

 \begin{figure*}[ht!]
    \centering
    \begin{subfigure}[b]{0.5\textwidth}
     \includegraphics[width=\textwidth]{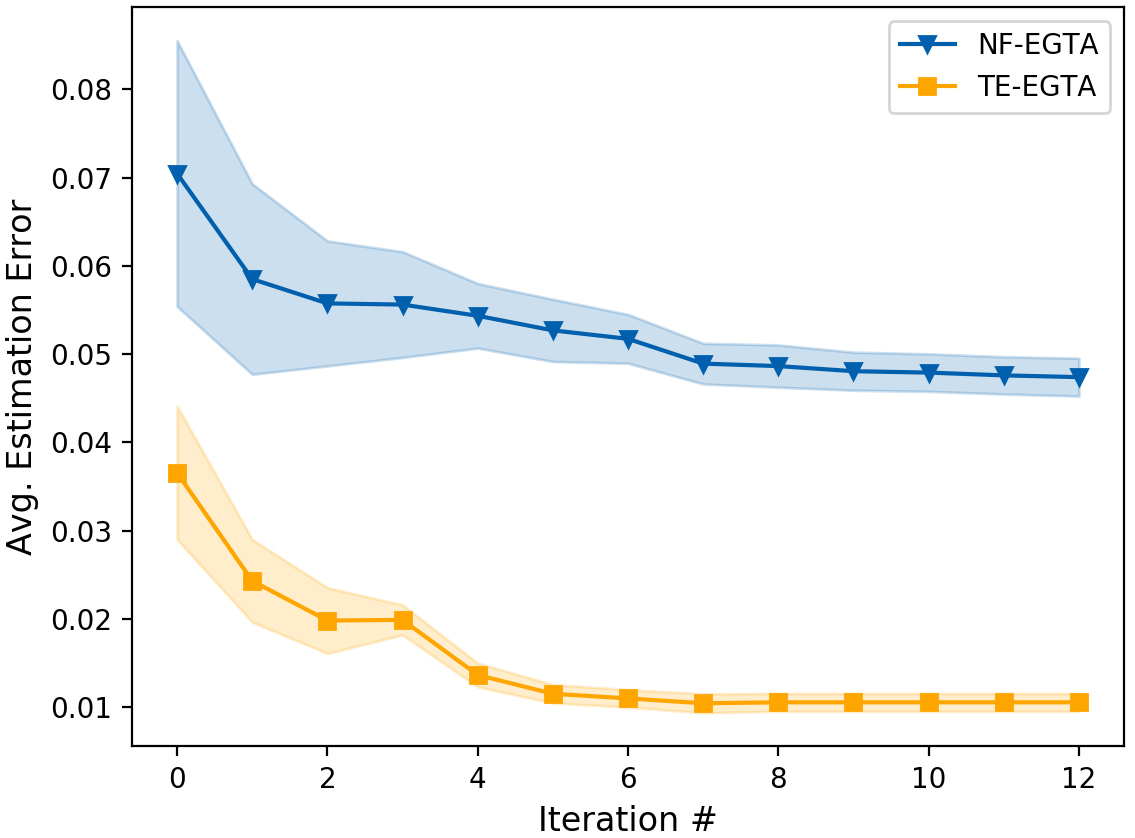}
    \caption{$m=200$ \label{fig:toy_err_200}}
    \end{subfigure}~
    \begin{subfigure}[b]{0.5\textwidth}
     \includegraphics[width=\textwidth]{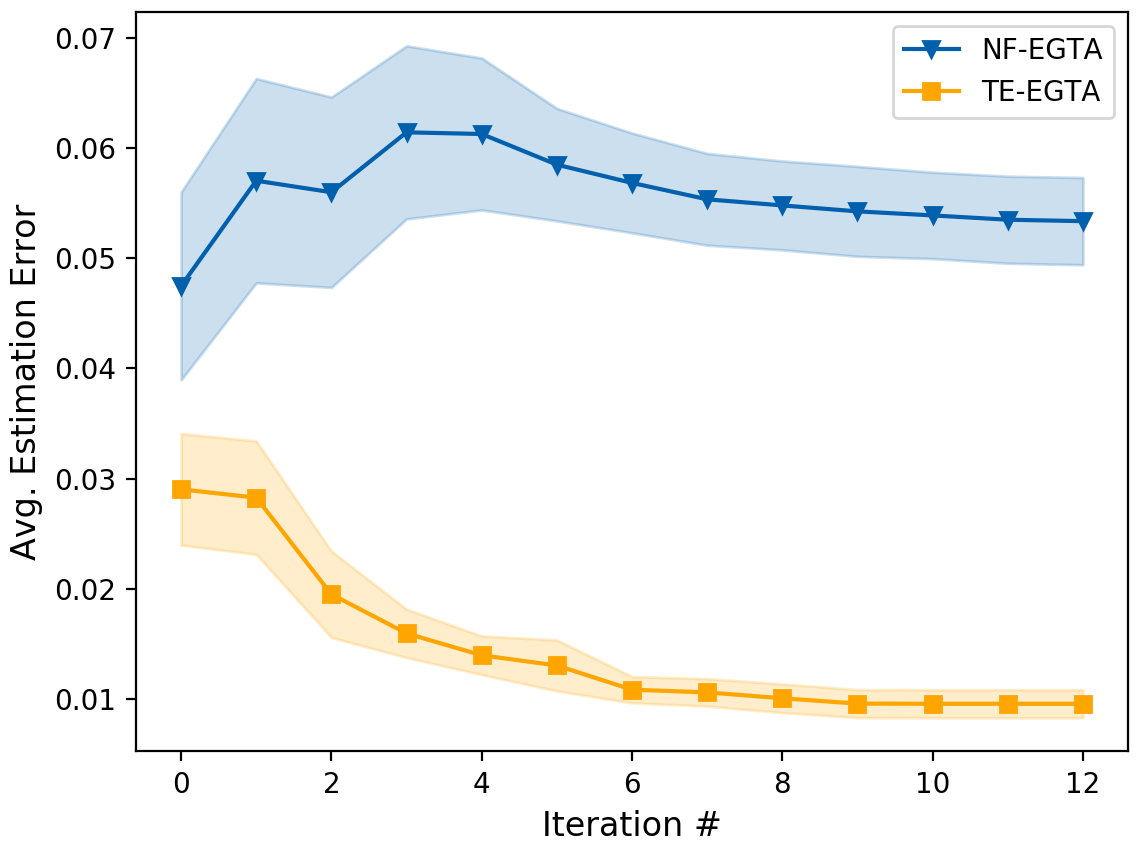}
    \caption{$m=100$ \label{fig:toy_err_100}}
    \end{subfigure}
    \caption{Comparing the average estimation error of the true EFG strategy payoffs over the course of EGTA's runtime for \tg, with $m=200$ game-play samples allotted for each strategy combination during simulation. Error bars represent the (estimated) standard error of the mean. \label{fig:est_err_200}}
\end{figure*}

\begin{figure*}[ht!]
    \centering
    \begin{subfigure}[b]{0.5\textwidth}
     \includegraphics[width=.97\textwidth]{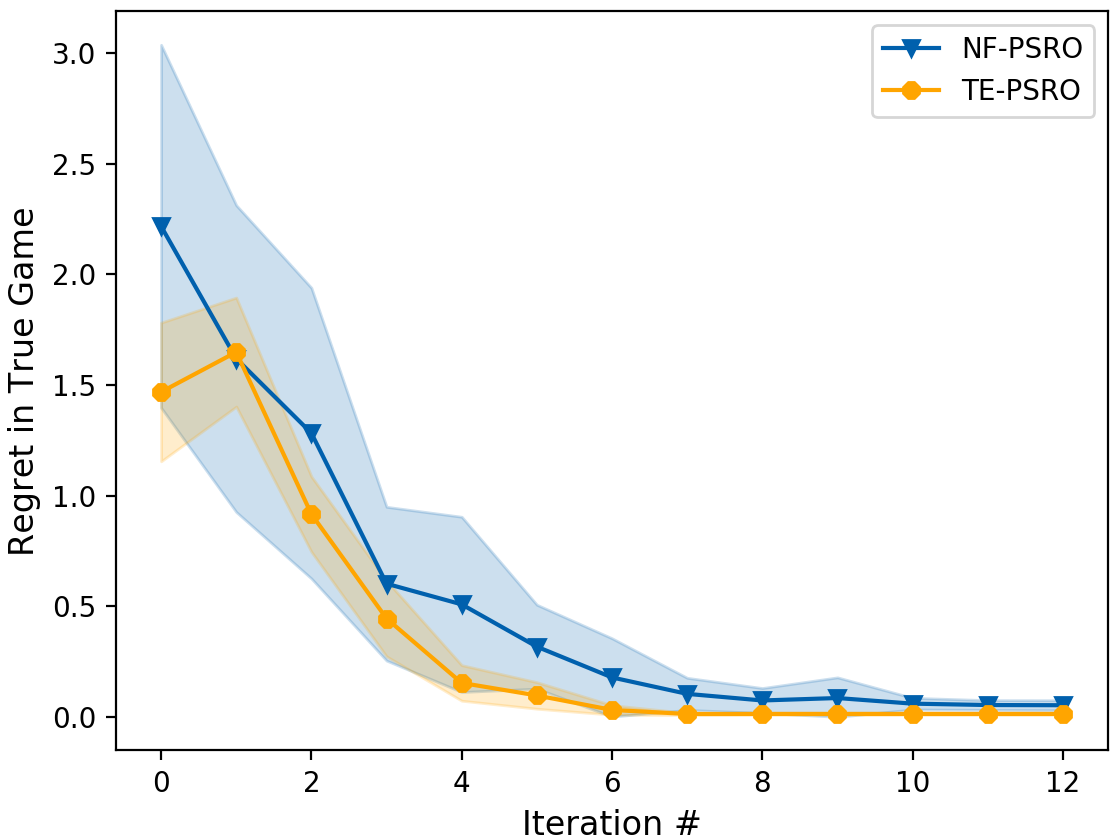}
    \caption{$m=200$ \label{fig:toy_reg_200}}
    \end{subfigure}~
    \begin{subfigure}[b]{0.5\textwidth}
     \includegraphics[width=\textwidth]{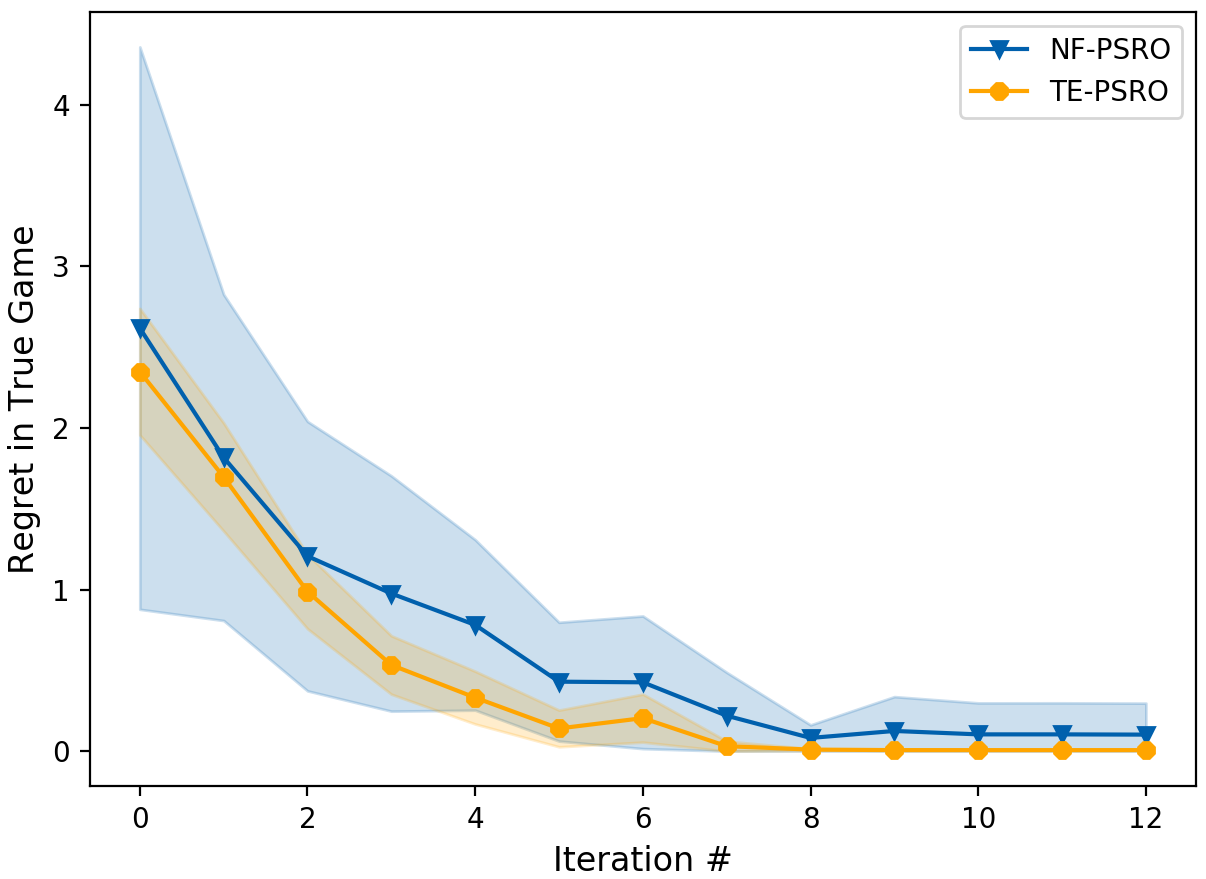}
    \caption{$m=100$  \label{fig:toy_reg_100}}
    \end{subfigure}
    \caption{Comparing the average regret of meta-strategy profiles over time for \tg, with $m$ game-play samples allotted for each strategy combination during simulation. Error bars represent the (estimated) standard error of the mean. \label{fig:reg_200}}
\end{figure*}



\begin{figure*}[ht!]
    \centering
    \includegraphics[width=0.6\textwidth]{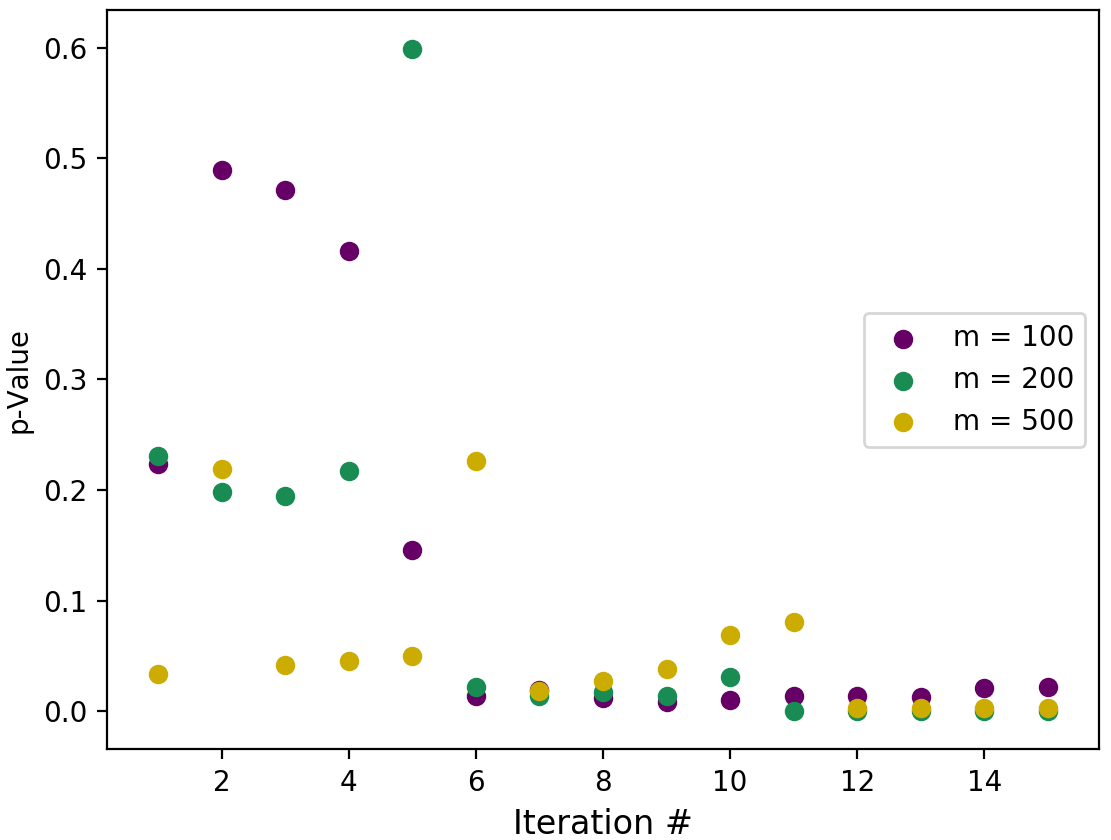}
    \caption{\tg \label{fig:toy_p_values_plot}}
    \caption{Reported p-values for the one-sided two-sample t-test performed on the results of \tg with null hypothesis $r_{EF} - r_{NF} \geq 0$ over each iteration of PSRO, for different values $m$ of allotted game-play samples per strategy combination during simulation. 
The null hypothesis is that the player regret resulting from TE-PSRO is at least as large as the regret resulting from NF-PSRO. 
}
    \label{fig:p_values_plots}
\end{figure*}
\end{document}